\documentclass[final,12pt]{elsarticle}

\usepackage{amssymb}
\usepackage[cmex10]{amsmath}
\usepackage{amsfonts}
\usepackage{amssymb}
\usepackage{amsthm}
\usepackage[linesnumbered,ruled,vlined]{algorithm2e}
\usepackage{graphicx}
\usepackage{caption}
\usepackage{subcaption}
\usepackage{multirow}
\usepackage{hyperref}
\usepackage{breakurl}

\begin{document}

\begin{frontmatter}



\title{Multilevel Hierarchical Kernel Spectral Clustering for Real-Life Large Scale Complex Networks}


\author{Raghvendra Mall, Rocco Langone and Johan A.K. Suykens}

\address{Department of Electrical Engineering, KU Leuven, ESAT-STADIUS, \\
 Kasteelpark Arenberg,10 B-3001 Leuven, Belgium\\ \{raghvendra.mall,rocco.langone,johan.suykens\}@esat.kuleuven.be}

\begin{abstract}
Kernel spectral clustering corresponds to a weighted kernel principal component analysis problem in a constrained optimization framework. The primal formulation leads to an eigen-decomposition of a centered Laplacian matrix at the dual level. The dual formulation allows to build a model on a representative subgraph of the large scale network in the training phase and the model parameters are estimated in the validation stage. The KSC model has a powerful out-of-sample extension property which allows cluster affiliation for the unseen nodes of the big data network. In this paper we exploit the structure of the projections in the eigenspace during the validation stage to automatically determine a set of increasing distance thresholds. We use these distance thresholds in the test phase to obtain multiple levels of hierarchy for the large scale network. The hierarchical structure in the network is determined in a bottom-up fashion. We empirically showcase that real-world networks have multilevel hierarchical organization which cannot be detected efficiently by several state-of-the-art large scale hierarchical community detection techniques like the Louvain, OSLOM and Infomap methods. We show a major advantage our proposed approach i.e. the ability to locate good quality clusters at both the coarser and finer levels of hierarchy using internal cluster quality metrics on $7$ real-life networks.     
\end{abstract}

\begin{keyword}
Hierarchical Community Detection \sep Kernel Spectral Clustering \sep Out-of-sample extensions



\end{keyword}

\end{frontmatter}


\section{Introduction}\label{sec:sec1}
Large scale complex networks are ubiquitous in the modern era. Their presence spans a wide range of domains including social networks, trust networks, biological networks, collaboration networks, financial networks etc. A complex network can be represented as a graph $G=(V,E)$ where $V$ represent the vertices or nodes and $E$ represents the edges or interaction between these nodes in this network. Many real-life complex networks are scale-free \cite{barabasi}, follow the power law \cite{clauset} and exhibit community like structure. By community like structure one means that nodes within one community are densely connected to each other and sparsely connected to nodes outside that community. The large scale network consists of several such communities. This problem of community detection in graphs has received wide attention from several perspectives \cite{girvan:newman, fortunato, danon:arenas, cnm, rosvall:bergstrom, Schaffer, lanchichinetti:fortunato, ng:jordan, shi:malik, luxburg,chung, manor:peronas}. 

The community structure exhibited by the real world complex networks often have an inherent hierarchical organization. This suggests that there should be multiple levels of hierarchy in these real-life networks with good quality clusters at each level. In other words, there exist meaningful communities at coarser as well as refined levels of granularity in this multilevel hierarchical system of the real-life complex networks. 

A state-of-the-art hierarchical community detection technique for large scale networks is the Louvain method \cite{blondel}. It uses a popular quality function namely \emph{modularity} (Q) \cite{girvan:newman,cnm,danon:arenas,newman} for locating modular structures in the network in a hierarchical fashion. Modularity  measures the difference between a given partition of a network and the expectation of the same partition for a random network. By optimizing modularity, they obtain the modular structures in the network. However, it suffers from a drawback namely the resolution limit problem \cite{fortunato:berthelemy,kumpula,good}. The issue of resolution limit arises because the optimization of modularity beyond a certain resolution is unable to identify modules even as distinct as cliques which are completely disconnected from the rest of the network. This is because modularity fixes a global resolution to identify modules which works for some networks but not others.

Recently the authors of \cite{lanchichinetti} show that methods trying to use variants of modularity to overcome the resolution limit problem, still suffer from the resolution limit. They propose an alternative algorithm namely OSLOM \cite{lanchichinetti:radicchi} to avoid the issue of resolution. However, in our experiments we observe that OSLOM works well for benchmark synthetic networks \cite{fortunato} but in case of real-life networks it is unable to detect quality clusters at coarser levels of granularity. We also evaluate another state-of-the-art hierarchical community detection technique called the Infomap method \cite{rosvall:bergstrom}. The Infomap method uses an information theoretic approach to hierarchical community detection. It uses the probability flow of random walks as a substitute for information flow in real-life networks. It then fragments the network into modules by compressing a description of the probability flow.      

Spectral clustering methods \cite{ng:jordan,shi:malik,luxburg,chung,manor:peronas} belong to the family of unsupervised learning algorithms where clustering information is obtained by the eigen-decomposition of the Laplacian matrix derived from the affinity matrix ($S$) for the given data. A drawback of these methods is the construction of the large affinity matrix for the entire data which limits the feasibility of the approach to small sized data. To overcome this problem, a kernel spectral clustering (KSC) formulation based on weighted kernel principal component analysis (kPCA) in a primal-dual framework was proposed in \cite{alzate:suykens}. The weighted kPCA problem is formulated in the primal in the context of least squares support vector machines \cite{johan,mall4,mall5,mall6} which results in eigen-decomposition of a centered Laplacian matrix in the dual. As a result, a clustering model is obtained in the dual. This model is build on a subset of the original data and has a powerful out-of-sample extension property. The out-of-sample extensions property allows cluster affiliation for unseen data. KSC is a state-of-the-art clustering technique which has been used for various applications \cite{mall7,mall8}.   

The KSC method was applied for community detection in graphs by \cite{langone:alzate}. However, their subset and model selection approach was computationally expensive and memory inefficient. Recently, the KSC method was extended for big data networks in \cite{mall1}. The method works by building a model on a representative subgraph of the large scale network. This subgraph is obtained by the fast and unique representative subset (FURS) selection technique as proposed in \cite{mall2}. During the model selection stage, the model parameters are estimated along with determining the number of clusters $k$ in the network. A self-tuned KSC model for big data networks was proposed in \cite{mall3}. The major advantage of the KSC method is that it creates a model which has a powerful out-of-sample extensions property. Using this property, we can infer community affiliation for unseen nodes of the whole network. 

In \cite{alzate}, the authors used multiple scales of the kernel parameter $\sigma$ to determine the hierarchical structure in the data using KSC approach. However, in this approach the clustering model is trained for different values of $(k,\sigma)$ and evaluated for the entire dataset using the out-of-sample extension property. Then, a map is created to match the clusters at two levels of hierarchy. As stated by the authors in \cite{alzate}, during a merge there might be some data points of the merging clusters that go into a non-merging cluster which is then forced to join the merging cluster of the majority. In this paper, we overcome this problem and generate a natural hierarchical organization of the large scale network in an agglomerative fashion.  

The purpose of hierarchical community detection is to automatically locate multiple levels of granularity in the network with meaningful clusters at each level. The KSC method has been used effectively to obtain flat partitioning in real-world networks \cite{langone:alzate, mall1, mall3}. In this paper, we exploit the structure of the eigen-projections derived from the KSC model. The projections of the validation set nodes in the eigenspace is used to create an iterative set of affinity matrices resulting in a set of increasing distance thresholds (${\cal T}$). Since the validation set of nodes is a representative subset of the large scale network \cite{mall2}, we use these distance thresholds ($t_{i} \in {\cal T}$) on the projections of the entire network obtained as a result of the out-of-sample extension property of the KSC model. These distance thresholds, when applied in an iterative manner, provide a multilevel hierarchical organization for the entire network in a bottom-up fashion. We show that our proposed approach is able to discover good quality coarse as well as refined clusters for real-life networks.

\begin{figure}[!ht]
	\centering
	\includegraphics[width=\textwidth]{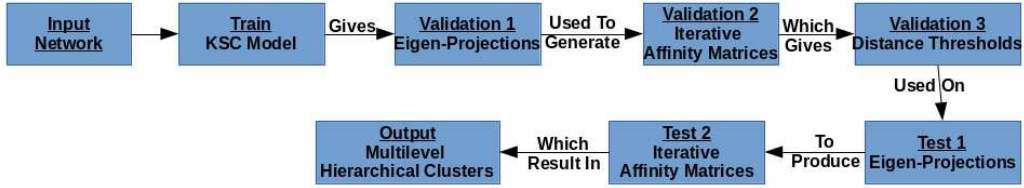}
	\caption{Steps undertaken by the MH-KSC algorithm}
	\label{fig:fig1}
\end{figure}

\begin{figure}[!ht]
	\centering
	\begin{subfigure}{\textwidth}
	{
		\centering
		\includegraphics[width=\textwidth]{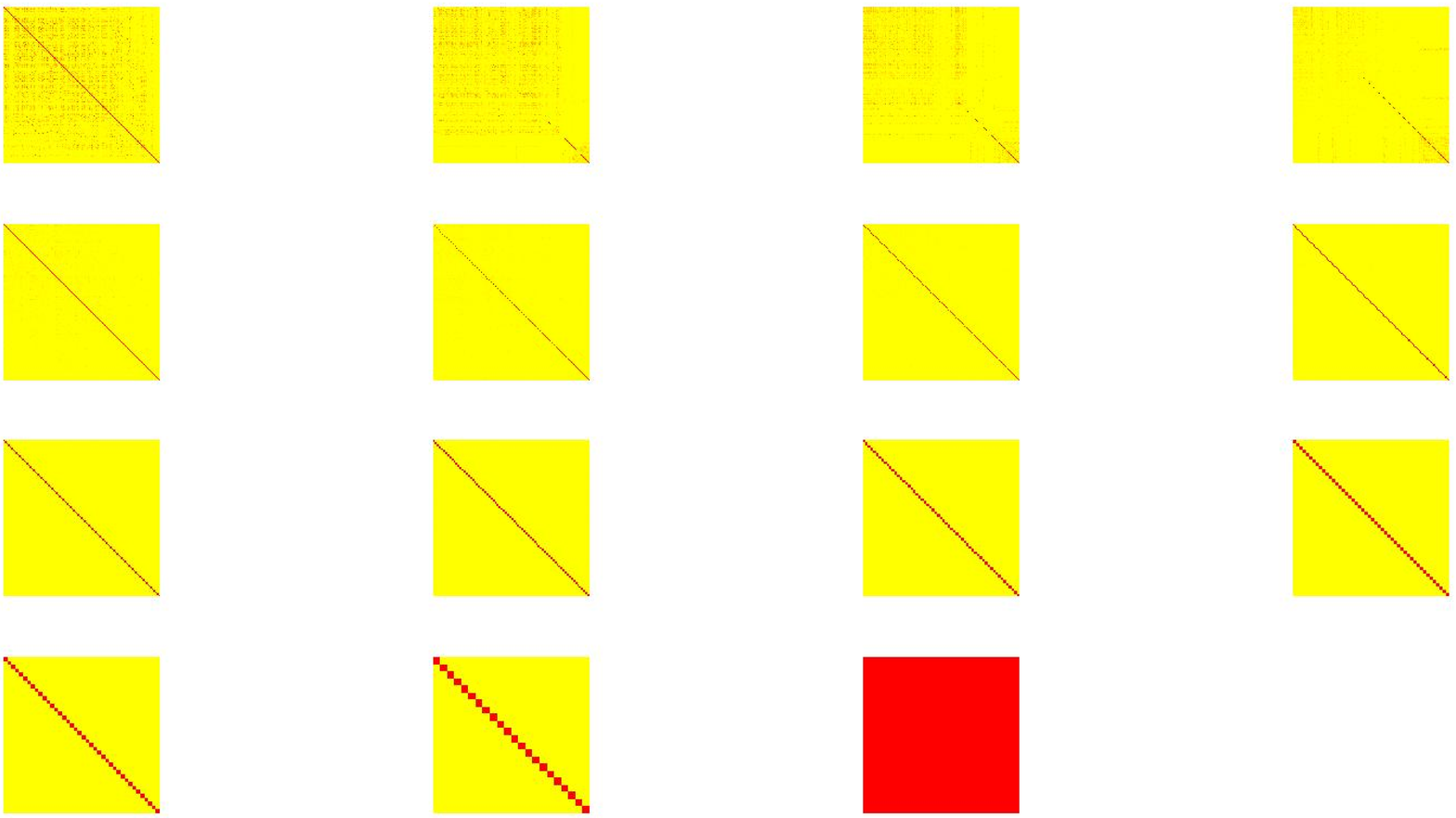}
		\caption{Affinity matrix created at different levels of hierarchy in left to right order. The number of block-diagonals in each subgraph represents  $k$ at that level of hierarchy.}
		\label{fig:subfig1}
	}
	\end{subfigure}
	\begin{subfigure}{\textwidth}
	{
		\centering
		\includegraphics[width=0.4\textwidth]{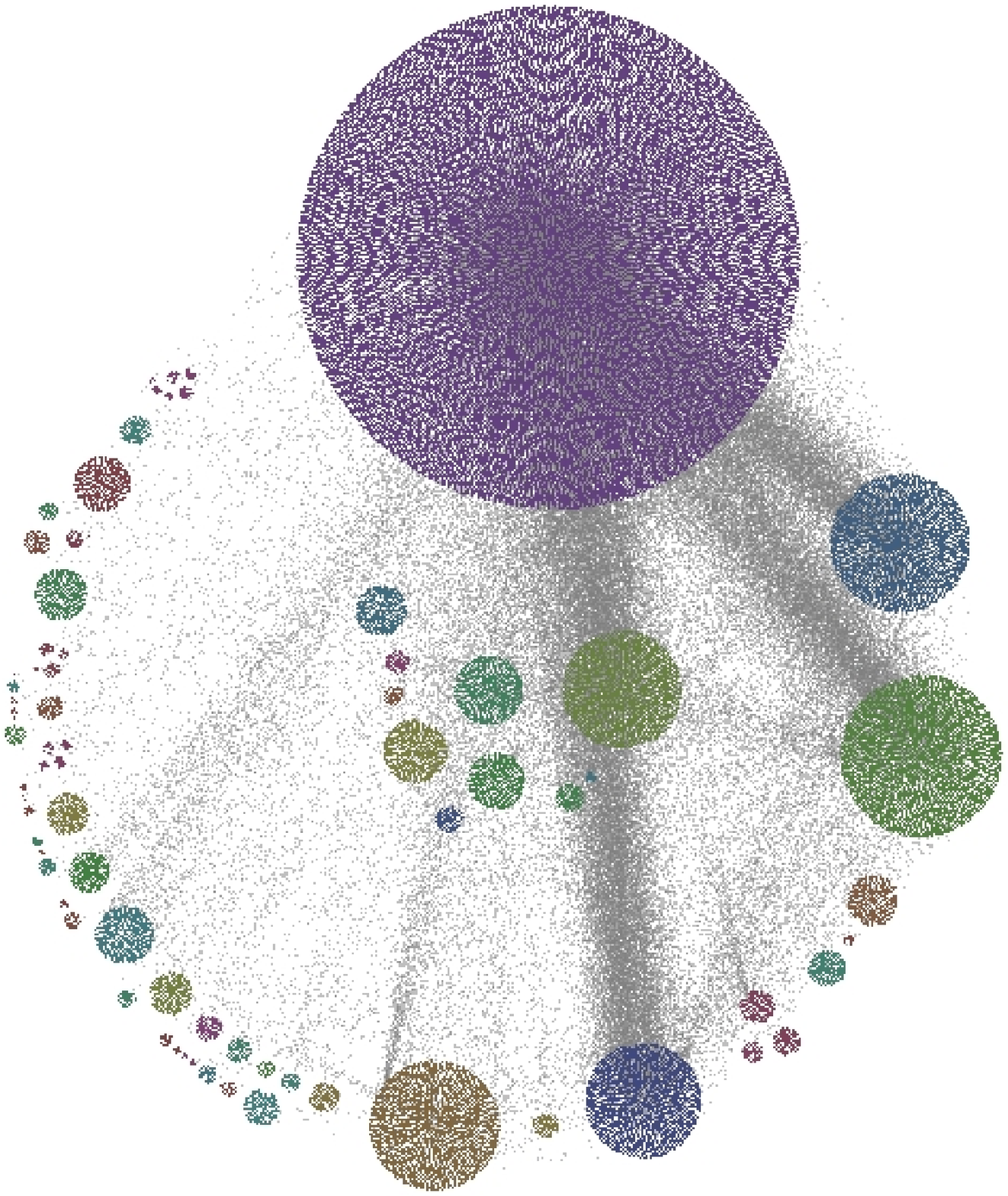}
		\caption{Result of MH-KSC algorithm on Enron dataset. Circles which have the same colour are part of the same cluster at the coarsest level of hierarchy. We depict clusters at $2$ different levels of hierarchy using the toolbox provided in \cite{lanchichinetti:radicchi}.}
		\label{fig:subfig2}
	} 
	\end{subfigure}
	\caption{Result of proposed MH-KSC approach on the Enron network}\label{fig:fig2}	
\end{figure}

There are some methods that optimize weighted graph cut objectives \cite{dhillon, metis, kushnir} to provide multilevel clustering for the large scale network. However, these methods suffer from the problem of determining the right value of $k$ which is user defined. In real-world networks the value of $k$ is not known beforehand. So in our experiments, we evaluate the proposed multilevel hierarchical kernel spectral clustering (MH-KSC) algorithm against the Louvain, Infomap and OSLOM methods. These methods automatically determine the number of clusters ($k$) at each level of hierarchy. Figure \ref{fig:fig1} provides an overview of steps involved in the MH-KSC algorithm and Figure \ref{fig:fig2} depicts the result of our proposed MH-KSC approach on email network (Enron).

In all our experiments we consider unweighted and undirected networks. All the experiments were performed on a machine with 12Gb RAM, 2.4 GHz Intel Xeon processor. The maximum size of the kernel matrix that is allowed to be stored in the memory of our PC is $10,000\times 10,000$. Thus, the maximum cardinality of our training and validation sets can be $10,000$. We use $15\%$ of the total nodes as size of training and validation set (if less than $10,000$) based on experimental findings in \cite{leskovec}. We make use of the procedure provided in \cite{mall1} to divide the data into chunks in order to extend our proposed approach to large scale networks. There are several steps in the proposed methodology which can be implemented on a distributed environment. They are described in detail in Section \ref{sec:sec3:subsec4}. 

Section \ref{sec:sec2} provides a brief description of the KSC method. Section \ref{sec:sec3} details the proposed multilevel hierarchical kernel spectral clustering algorithm. The experiments, their results and analysis are described in Section \ref{sec:sec4}. We conclude the paper with Section \ref{sec:sec5}. 
%%%%%%%%%%%%%%%%%%%%%%%%%%%%%%%%%%%%%%%%%5
\section{Kernel Spectral Clustering (KSC) method}\label{sec:sec2}
We first summarize the notations used in the paper.
\subsection{Notations}
\begin{enumerate}
\item A graph is mathematically represented as $G = (V,E)$ where $V$ represents the set of nodes and $E \subseteq V\times V$ represents the set of edges in the network. Physically, the nodes represent the entities in the network and the edges represent the relationship between these entities.
\item The cardinality of the set $V$ is denoted as $N$.
\item The training, validation and test set of nodes is given by $V_{tr}$, $V_{valid}$ and $V_{test}$ respectively.
\item The cardinality of the training, validation and test set is given $N_{tr}$, $N_{valid}$, $N_{test}$.
\item The adjacency list corresponding to each vertex $v_{i} \in V$ is given by $x_{i} = A(:,i)$.
\item $maxk$ is the maximum number of eigenvectors that we want to evaluate. 
\item $K(\cdot,\cdot)$ represents the positive definite kernel function.
\item The matrix $S$ represents the affinity or similarity matrix. 
\item $P$ represents the latent variable matrix containing the eigen-projections.
\item $h$ represents the $h^{th}$ level of hierarchy and $maxh$ stands for the coarsest level of hierarchy.
\item Set $C$ comprises multilevel hierarchical clustering information.
\end{enumerate}
%%%%%%%%%%%%%%%%%%%%%%%%%%%%%%%%%%%%%%%%%%%%%%%%%%%%
\subsection{KSC methodology}
Given a graph $G$, we perform the FURS selection \cite{mall2} technique to obtain training and validation set of nodes $V_{tr}$ and $V_{valid}$. For $V_{tr}$ training nodes the dataset is given by $\mathcal{D} = \{x_{i}\}_{i=1}^{N_{tr}}$, $x_{i} \in \mathbb{R}^{N}$. The adjacency list $x_{i}$ can efficiently be stored into memory as real-world networks are highly sparse and have limited connections for each node $v_{i}$. 

Given $\mathcal{D}$ and $maxk$, the primal formulation of the weighted kernel PCA \cite{alzate:suykens} is given by:
\begin{equation}\label{eq:eq1}
\begin{aligned}
& \underset{w^{(l)},e^{(l)},b_{l}}{\text{min}}
& & \frac{1}{2} \sum\limits_{l=1}^{maxk-1} {w^{(l)}}^{\intercal}w^{(l)} - \frac{1}{2N_{tr}}\sum\limits_{l=1}^{maxk-1} \gamma_{l}{e^{(l)}}^{\intercal}D_{\Omega}^{-1}e^{(l)} \\
& \text{such that}
& & e^{(l)} =  \Phi w^{(l)} + b_{l}1_{N_{tr}}, l=1, \ldots, maxk-1,
\end{aligned}
\end{equation} 
where $e^{(l)} = [{e}_{1}^{(l)}, \ldots , {e}_{N_{tr}}^{(l)}]^{\intercal}$ are the projections onto the eigenspace, $l=1,\ldots, maxk$-$1$ indicates the number of score variables required to encode the $maxk$ clusters. However, it was shown in \cite{mall3} that we can discover more than $maxk$ communities using these $maxk$-$1$ score variables. $D_{\Omega}^{-1} \in \mathbb{R}^{N_{tr}\times N_{tr}}$ is the inverse of the degree matrix associated to the kernel matrix $\Omega$ with $\Omega_{ij} = K(x_{i},x_{j}) = \phi(x_{i})^\intercal \phi(x_{j})$. $\Phi$ is the $N_{tr}\times d_{h}$ feature matrix such that $\Phi = [\phi(x_{1})^{\intercal};\ldots;\phi(x_{N_{tr}})^{\intercal}]$ and $\gamma_{l} \in \mathbb{R}^{+}$ is the regularization constant. We note that $N_{tr}\ll N$ i.e. the number of nodes in the training set is much less than the total number of nodes in the large scale network. 

The kernel matrix $\Omega$ is constructed by calculating the similarity between the adjacency list of each pair of nodes in the training set. Each element of $\Omega$, defined as $\Omega_{ij} = \frac{x_{i}^{\intercal}x_{j}}{\| x_{i}\|\| x_{j} \|}$ is calculated by estimating the cosine similarity between the adjacency lists $x_{i}$ and $x_{j}$ using notions of set intersection and union. This corresponds to using a normalized linear kernel function $K(x,z) = \frac{x^{\intercal}z}{\|x\| \|z\|}$ \cite{johan}.

The primal clustering model is then represented by:
\begin{equation}\label{eq:eq2}
e_{i}^{(l)} = {w^{(l)}}^{\intercal}\phi(x_{i}) + b_{l}, i=1,\ldots , N_{tr},
\end{equation} 
where $\phi: \mathbb{R}^{N} \rightarrow \mathbb{R}^{d_{h}}$ is the feature map i.e. a mapping to high-dimensional feature space $d_{h}$ and $b_{l}$ are the bias terms, $l=1,\ldots , maxk$-$1$. For large scale networks we can utilize the explicit expression of the underlying feature map as shown in \cite{mall1} and set $d_{h}=N$. The dual problem corresponding to this primal formulation is given by:
\begin{equation}\label{eq:eq3}
D_{\Omega}^{-1}M_{D}\Omega \alpha^{(l)} = \lambda_{l}\alpha^{(l)},
\end{equation} 
where $M_{D}$ is the centering matrix which is defined as $M_{D} = {\rm I}_{N_{tr}} -  (\frac{(1_{N_{tr}}1_{N_{tr}}^{\intercal}D_{\Omega}^{-1})}{1_{N_{tr}}^{\intercal}D_{\Omega}^{-1}1_{N_{tr}}})$. The $\alpha^{(l)}$ are the dual variables and the kernel function $K: \mathbb{R}^{N}\times \mathbb{R}^{N} \rightarrow \mathbb{R}$ plays the role of similarity function. The dual predictive model is:
\begin{equation}\label{eq:eq4}
\hat{e}^{(l)}(x) = \sum_{i=1}^{N_{tr}} \alpha_{i}^{(l)} K(x,x_{i}) + b_{l},
\end{equation}
which provides clustering inference for the adjacency list $x$ corresponding to the validation or test node $v$. 
%%%%%%%%%%%%%%%%%%%%%%%%%%%%%%%%%%%%%%%%%%%%%%%%%%%%%%%%%%%%%%
\section{Multilevel Hierarchical KSC}\label{sec:sec3}
We use the predictive KSC model in the dual to get the latent variable matrix for the validation set $V_{valid}$ represented as $P_{valid} = [e_{1},\ldots,e_{N_{valid}}]^\intercal$ and the test set $V_{test}$ (entire network) denoted by $P_{test}$. In \cite{mall3} the authors create an affinity matrix $S_{valid}$ using the latent variable matrix $P_{valid}$ which is a $N_{valid} \times (maxk$-$1)$ matrix, as:
\begin{equation}\label{eq:eq5}
S_{valid}(i,j) = CosDist(e_{i},e_{j}) = 1 - \cos(e_{i},e_{j}) = 1 - \frac{ e_{i}^\intercal e_{j}}{\|e_{i} \| \| e_{j} \|},
\end{equation}
where $CosDist(\cdot,\cdot)$ function calculates the cosine distance between $2$ vectors and takes values between $[0,2]$. Nodes which belong to the same community will have $CosDist(e_{i},e_{j})$ closer to $0$, $\forall i,j$ in the same cluster. It was shown in \cite{mall3} that a rotation of the $S_{valid}$ matrix has a block diagonal structure. This block diagonal structure was used to identify the ideal number of clusters $k$ in the network using the concept of entropy and balanced clusters. 
%%%%%%%%%%%%%%%%%%%%%%%%%%%%%%%%%%%%
\subsection{Determining the Distance Thresholds}\label{sec:sec3:subsec1}
We propose an iterative bottom-up approach on the validation set to determine the set of distance thresholds ${\cal T}$. In our approach, we refer to the affinity matrix at the ground level of hierarchy as $S_{valid}^{(0)}$. The $S_{valid}^{(0)}$ matrix is obtained by calculating the $CosDist(\cdot,\cdot)$ between each element of the latent variable matrix $P_{valid}$ as mentioned earlier. After several empirical evaluations, we observe that distance threshold at level $0$ of hierarchy can be set to values between $[0.1,0.2]$. In our experiments we set $t^{(0)}=0.15$.  This allows to make the approach tractable to large scale networks which will be explained in section \ref{sec:sec3:subsec2}.

We then use a greedy approach to select the validation node with maximum number of similar nodes in the latent space i.e. we select the projection $e_{i}$ which has a maximum number of projections $e_{j}$ satisfying $S_{valid}^{(0)}(i,j) < t^{(0)}$. We put the indices of these nodes in $C^{(0)}_{1}$ representing the $1^{st}$ cluster at level $0$ of hierarchy. We then remove these nodes and corresponding entries from $S_{valid}^{(0)}$ to obtain a reduced matrix. This process is repeated iteratively until $S_{valid}^{(0)}$ becomes empty. Thus, we obtain the set $C^{(0)} = \{C^{(0)}_{1},\ldots,C^{(0)}_{q}\}$ where $q$ is the total number of clusters at ground level of hierarchy. The set $C^{(0)}$ has communities along with the indices of the nodes in these communities. 

To obtain the clusters at the next level of hierarchy we treat the communities at the previous levels as nodes. We then calculate the average cosine distance between these nodes using the information present in them. At each level $h$ of hierarchy we create a new affinity matrix as:
\begin{equation}\label{eq:eq6}
S_{valid}^{(h)}(i,j) = \frac{\sum_{m \in C^{(h-1)}_{i}} \sum_{l \in C^{(h-1)}_{j}} S_{valid}^{(h-1)}(m,l)}{|C^{(h-1)}_{i}|\times |C^{(h-1)}_{j}|},
\end{equation}
where $|\cdot|$ represents the cardinality of the set. In order to determine the threshold at level $h$ of hierarchy, we estimate the minimum cosine distance between each individual cluster and the other clusters (not considering itself). Then, we select the mean of these values as the new threshold for that level to combine clusters. This makes the approach different from the classical single-link clustering where we combine two clusters which are closest to each other at a given level of hierarchy and the average-link agglomerative clustering where we combine based on the average distance between all the clusters.

The reason for using mean of these minimum cosine distance values as the new threshold is that if we consider the minimum of all the distance values then there is a risk of only combining $2$ clusters at that level. However, it is desirable to combine multiple sets of different clusters. Thus, the new threshold $t^{(h)}$ at level $h$ is set as: 
\begin{equation}\label{eq:eq7}
t^{(h)} = \text{mean}(\text{min}_{j}(S_{valid}^{(h)}(i,j))), i \neq j.
\end{equation}

We use this process iteratively till we reach the coarsest level of hierarchy where we have $1$ cluster containing all the nodes. As a consequence we obtain the hierarchical clustering ${\cal C} = \{C^{(0)},\ldots,C^{(maxh)}\}$ automatically. As we move from one level of hierarchy to another the value of distance threshold increases since we are merging large clusters at coarser levels of hierarchy. We finally end up with a set of increasing distance thresholds ${\cal T} = \{t^{(0)},\ldots,t^{(maxh)}\}$.
%%%%%%%%%%%%%%%%%%%%%%%%%%%%%%%%%%%
\subsection{Requirements for Feasibility to Large Scale Networks}\label{sec:sec3:subsec2}
The whole large scale network is used as test set. The latent variable matrix for the test set is obtained by out-of-sample extensions of the predictive KSC model and defined as $P_{test} = [e_{1},\ldots,e_{N_{test}}]^\intercal$. Since we use the entire network as test set, therefore, $N_{test}=N$. The $P_{test}$ matrix is a $N \times (maxk$-$1)$ dimensional matrix. So, we can store this $P_{test}$ matrix in memory but cannot create an affinity matrix of size $N \times N$ due to memory constraints.

To make the approach feasible to large scale network we put a condition that the maximum size of a cluster at ground level cannot exceed $10,000$ (depending on the available computer memory) and the maximum number of clusters allowed at the ground level is $10,000$. This limits the size of the affinity matrix at that level of hierarchy to be less than $10,000 \times 10,000$. It also effects the choice of the  initial value of the distance threshold $t^{(0)}$. If we set $t^{(0)}$ too high ($\gg 0.2$) then majority of the nodes at the ground level in the test case will fall in one community resulting in one giant connected component. If we set the value of $t^{(0)}$ too low ($\ll 0.1$) then we will end up with lot of singleton clusters at the ground level in the test case. In our experiments, we observed that the interval any value between $[0.1,0.2]$ is good choice for the initial threshold value at level $0$ of hierarchy. To be consistent we chose $t^{(0)}=0.15$ for all the networks.
%%%%%%%%%%%%%%%%%%%%%%%%%%%%%%%%%%%%%%%%%%%%%%%%%%%%%
\subsection{Multilevel Hierarchical KSC for Test Nodes}\label{sec:sec3:subsec3}
The validation set is a representative subset of the whole network as shown in \cite{mall2}. Thus, the threshold set ${\cal T}$ can be used to obtain a hierarchical clustering for the entire network. To make the proposed approach self-tuned, we use $t^{(i)} > t^{(0)} > 0.15$, $i>0$, during the test phase. 

In order to prevent creating the affinity matrix for the large network we follow a greedy procedure. We select the projection of the first test node and calculate its similarity with the projections of all the test nodes. We then locate the indices ($j$) of those projections s.t. $CosDist(e_{1},e_{j}) < t^{(1)}$. If the total number of such indices is less than $10,000$ then we put them in cluster $C^{(1)}_{1}$ otherwise we select the first $10,000$ indices and place them in cluster $C^{(1)}_{1}$. This is due to the constraint that the size of a cluster ($C^{(1)}_{1}$) at ground level cannot exceed $10,000$. We then remove entries corresponding to those projections in $P_{test}$ to obtain a reduced matrix. We perform this procedure iteratively until $P_{test}$ is empty to obtain $C^{(1)} = \{C^{(1)}_{1},\ldots,C^{(1)}_{r}\}$ where $r$ is the total number of clusters at hierarchical level $1$. After the $1^{st}$ level, we use the same procedure that was for validation set i.e. creating an affinity matrix at each level using the cluster information along with the threshold set ${\cal T}$ to obtain the hierarchical structure in an agglomerative fashion. The cluster memberships are propagated iteratively from the $1^{st}$ level to the highest level of hierarchy. The multilevel hierarchical kernel spectral clustering (MH-KSC) method is described in Algorithm \ref{alg:algo1}.  
\begin{algorithm}[!ht]
\SetAlgoLined
	\KwData{Graph $G=(V,E)$ representing large scale network.}
	\KwResult{Multilevel Hierarchical Organization of the network.}
	Divide data into train,validation and test set, $V_{tr}$,$V_{valid}$,$V_{test}$.\\
	Construct dataset $\mathcal{D} = \{x_{i}\}_{i=1}^{N_{tr}}$, $x_{i} \in \mathbb{R}^{N}$ from training set $V_{tr}$. \\
	Perform KSC on $\mathcal{D}$ to obtain the predictive model as in (\ref{eq:eq4}).\\
	Obtain $P_{valid} = [e_{1},\ldots,e_{N_{valid}}]^\intercal$ using predictive model and $V_{valid}$.\\
	Construct $S^{(0)}_{valid}(i,j) = CosDist(e_{i},e_{j}) = 1 - \frac{ e_{i}^\intercal e_{j}}{\|e_{i} \| \| e_{j} \|}$, $\forall e_{i},e_{j} \in P_{valid}$.\\
	Begin validation stage with: $h=0$, $t^{(0)}=0.15$.\\
	$[C^{(0)},k] = GreedyMaxOrder(S^{(0)}_{valid},t^{(0)})$. \tcc{Algorithm \ref{alg:algo2}}
	Add $t^{(0)}$ to the set ${\cal T}$ and $C^{(0)}$ to the set ${\cal C}$.\\
	\While{$k>1$}{ $h:=h+1$. \\
		Create $S^{(h)}_{valid}$ using $S^{(h-1)}_{valid}$ and $C^{(h-1)}$ as shown in (\ref{eq:eq6}). \\
		Calculate $t^{(h)}$ using equation (\ref{eq:eq7}).\\	
		$[C^{(h)},k] = GreedyMaxOrder(S^{(h)}_{valid},t^{(h)})$.\\
		Add $t^{(h)}$ to the set ${\cal T}$ and $C^{(h)}$ to the set ${\cal C}$. \\
	} 
	\tcc{Iterative procedure to get the set ${\cal T}$.}
	Obtain $P_{test}$ like $P_{valid}$ and begin with: $h=1$, $t^{(1)} \in {\cal T}$.\\
	$[S^{(2)}_{test},C^{(1)},k] = GreedyFirstOrder(P_{test},t^{(1)})$. \tcc{Algorithm \ref{alg:algo3}}
	Add $C^{(1)}$ to the set ${\cal C}$.\\
	\ForEach{$t^{(h)} \in {\cal T}$, $h>1$}{	
		$[C^{(h)},k] = GreedyMaxOrder(S^{(h)}_{test},t^{(h)})$.\\
		Add $C^{(h)}$ to the set ${\cal C}$. \\
		Create $S^{(h+1)}_{test}$ using $S^{(h)}_{test}$ and $C^{(h)}$ as shown in (\ref{eq:eq6}). \\
	}
	Obtain the set ${\cal C}$ for test set and propagate cluster memberships iteratively from $1^{st}$ to coarsest level of hierarchy.
\caption{MH-KSC Algorithm}\label{alg:algo1}
\end{algorithm}
%%%%%%%%%%%%%%%%%%%%%%%%%%%%%%%%
\begin{algorithm}
\SetAlgoLined
	\KwData{Affinity matrix $S$ and threshold $t$.}
	\KwResult{Clustering information $C$ and number of clusters $k$.}
	$k=1$ and $totinst=0$.\\
	\While{$totinst \neq |S|$} {
		Find $i$ in range $(1,|S|)$ for which number of instances $j$, s.t. $S(i,j)<t$, $j=1,\ldots, |S|$, is maximum. \\
		Put indices of instance $i$ and all instances $j$, s.t. $S(i,j)<t$, to $C_{k}$. \\
		$k := k + 1$ and $totinst: = totinst+|C_{k}|$. \\
		Set all elements corresponding to the indices in $C_{k}$ to $\infty$ in $S$.\\
		Add $C_{k}$ to the set $C$.
	}
	$k:=k-1$.
\caption{$GreedyMaxOrder$ Algorithm}\label{alg:algo2}
\end{algorithm}
%%%%%%%%%%%%%%%%%%%%%%%%%%%%%%%%%%%%
\begin{algorithm}
\SetAlgoLined
	\KwData{Projection matrix $P_{test}$, threshold $t^{(1)}$.}
	\KwResult{Affinity matrix $S^{(2)}_{test}$, clustering information $C^{(1)}$ and  $k$.}
	$k=1$.\\
	\While{$|P_{test}| \neq 0$} {
		Select $1^{st}$ node and locate all nodes $j$ for which $CosDist(e_{1},e_{j})<t^{(1)}$. \\
		Put all these instances in $C^{(1)}_{k}$ and to set $C^{(1)}$.\\
		$k:=k+1$.\\
		Remove these instances from $P_{test}$ to have a reduced $P_{test}$.\\ 
	}
	\tcc{The affinity matrix ($S^{(1)}_{test}$) is not calculated as it would be unfeasible to store an $N \times N$ matrix in memory.}
	$k:=k-1$.\\
	\For {$i=1$ to $|C^{(1)}|$}{
		\For {$j=i+1$ to $|C^{(1)}|$}{
				Calculate $S^{(2)}_{test}(i,j)$ as the average $CosDist(\cdot,\cdot)$ between the eigen-projections of the instances in $C^{(1)}_{i}$ and $C^{(1)}_{j}$. 
				}
	}
	\tcc{Affinity Matrix calculated for the first time.}
\caption{$GreedyFirstOrder$ Algorithm}\label{alg:algo3}
\end{algorithm}
%%%%%%%%%%%%%%%%%%%%%%%%%%%%%%%%%%%%%%%%%%%%
\subsection{Time Complexity Analysis}\label{sec:sec3:subsec4}
The two steps in our proposed approach which require the maximum computation time are the out-of-sample extensions for the test set and the creation of the affinity matrix from the ground level clusters.

Since we use the entire network as test set the time required for out-of-sample extension is $O(N_{tr}\times N)$. Our greedy procedure to obtain the clustering information at the ground level $C^{(1)}$ requires $O(r\times N)$ computations where $r$ is the number of clusters at $1^{st}$ level of hierarchy for the test set. This is because for each cluster $C^{(1)}_{1} \in C^{(1)}$ we remove all the indices belonging in that cluster from the matrix $P_{test}$. As a result the size of $P_{test}$ decreases till it reduces to zero resulting in $O(r\times N)$ computations. 
The affinity matrix $S^{(1)}_{test}$ is a symmetric matrix so we only need to compute the upper of lower triangular matrix. The number of cluster-cluster similarities that we have to calculate is $\frac{r\times (r-1)}{2}$ where the size of each cluster at ground level can be maximum $10,000$. 

However, as shown in \cite{mall1}, we can perform the out-of-sample extensions in parallel on $n$ computers and rows of the affinity matrix can also be calculated in parallel thereby reducing the complexity by $\frac{1}{n}$.  
\section{Experiments}\label{sec:sec4}
We conducted experiments on $2$ synthetic datasets obtained from the toolkit in \cite{fortunato} and $7$ real-world networks obtained from \burl{http://snap.stanford.edu/data/index.html}. 
\subsection{Synthetic Network Experiments}
The synthetic networks are referred as $Net_{1}$ and $Net_{2}$ and have $2,000$ and $50,000$ nodes respectively. The ground truth for these $2$ benchmark networks are known at $2$ levels of hierarchy. These $2$ levels of hierarchy for these benchmark networks are obtained by using $2$ different mixing parameters i.e. $\mu_{1}$ and $\mu_{2}$ for macro and micro communities. We fixed $\mu_{1}=0.1$ and $\mu_{2}=0.2$ in our experiments. Since the ground truth is known beforehand, we evaluate the communities obtained by our proposed MH-KSC approach using an external quality metric like Adjusted Rand Index ($ARI$) and Variation of Information ($VI$) \cite{rabbany,mall9}. We also evaluate the cluster information using internal cluster quality metrics like Modualrity ($Q$) \cite{girvan:newman} and Cut-Conductance ($CC$) \cite{dhillon}. We compare MH-KSC with Louvain, Infomap and OSLOM. 

\begin{figure}[!ht]
	\centering
	\begin{subfigure}{\textwidth}
	{
		\centering
		\includegraphics[width=0.7\textwidth]{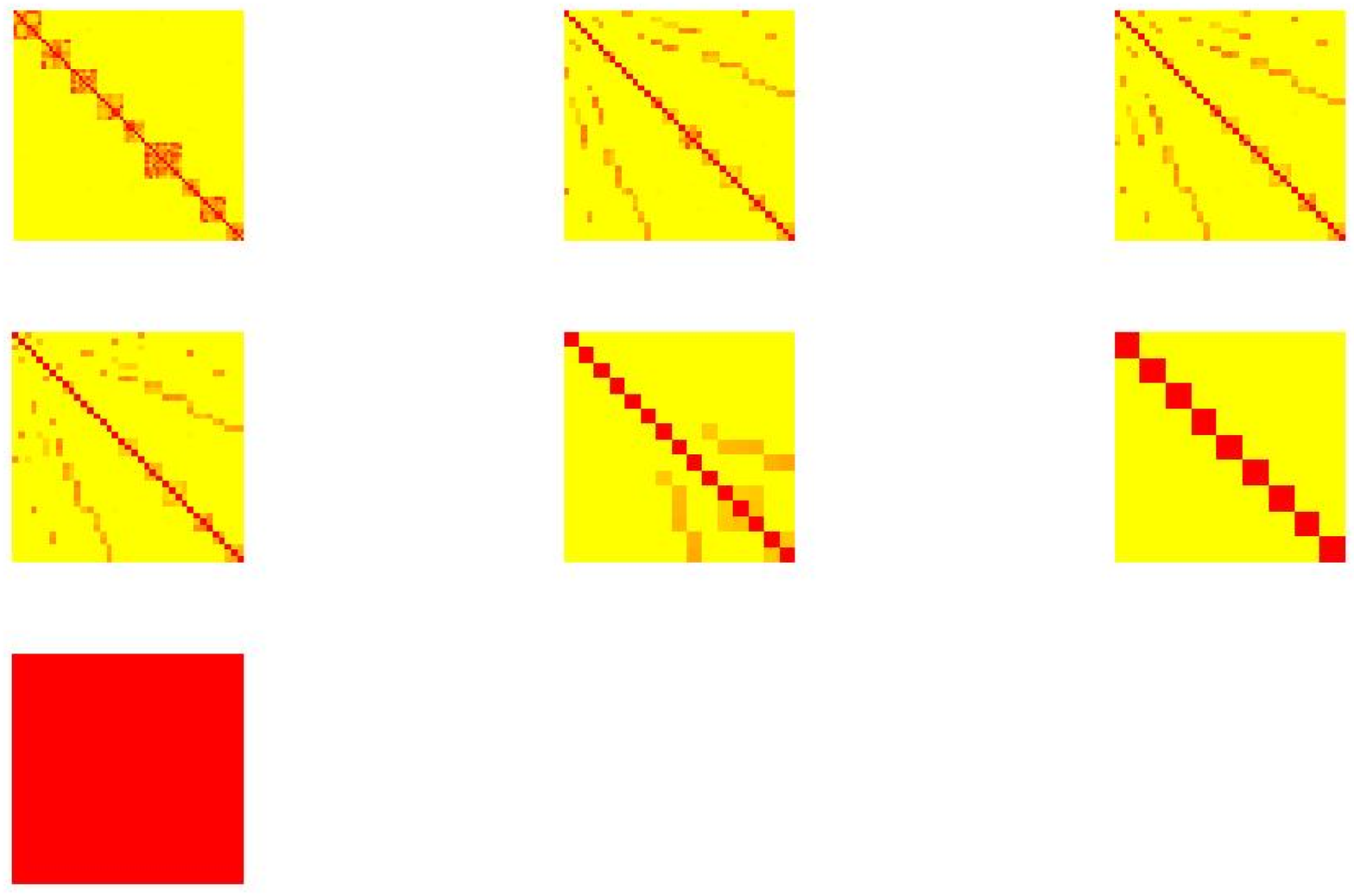}
		\caption{Affinity matrices created at different levels of hierarchy for $Net_{1}$ network. The number of block-diagonals in each subgraph represents  $k$ at that level of hierarchy.}
		\label{fig:subfig3}
	}
	\end{subfigure}
	\begin{subfigure}{\textwidth}
	{
		\centering
		\includegraphics[width=0.47\textwidth]{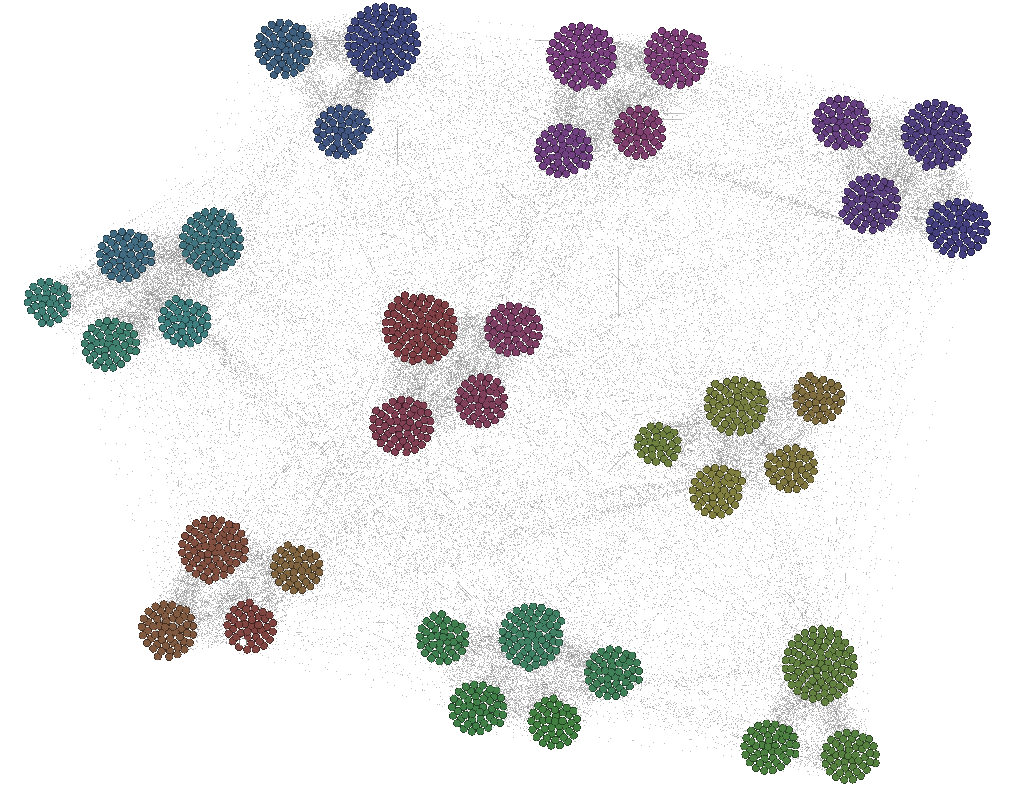}
		\includegraphics[width=0.47\textwidth]{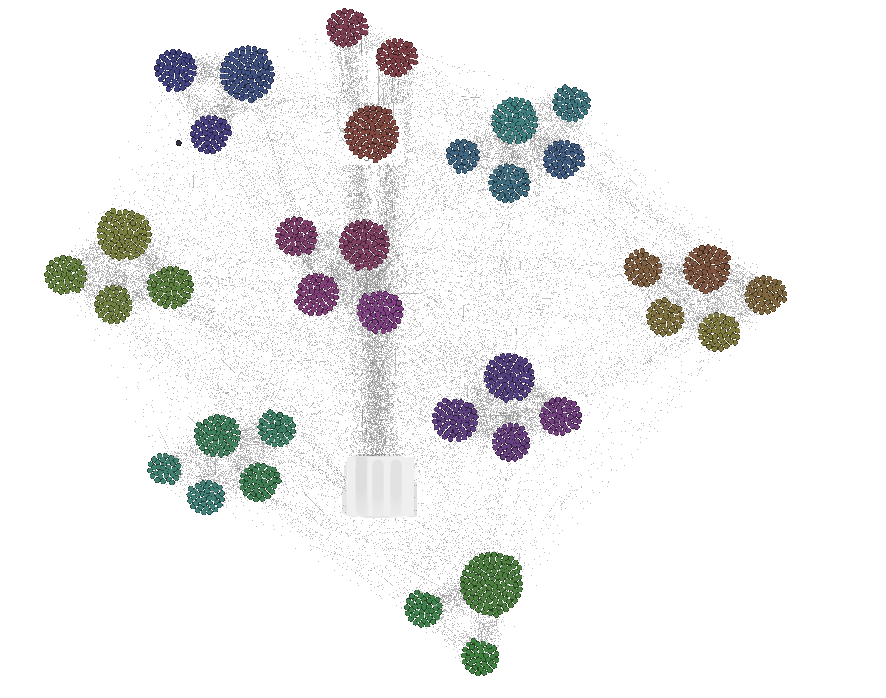}
		\caption{Original hierarchical network (left) and estimated hierarchical network (right) for synthetic network with $10,000$ nodes. The orientation and position of the communities might vary in the two plots. Both plots have $3$ clusters with $5$ micro communities, $4$ clusters with $4$ micro communities and $2$ clusters with $3$ micro communities.}
		\label{fig:subfig4}
	} 
	\end{subfigure}
	\caption{Result of MH-KSC algorithm on benchmark $Net_{1}$ network.}\label{fig:fig3}
\end{figure}
%%%%%%%%%%%%%%%%%%%%%%%%%%%%%%%%%%%%%%%%%%
\begin{figure}[!ht]
	\centering
	\begin{subfigure}{0.8\textwidth}
	{
		\centering
		\includegraphics[width=\textwidth]{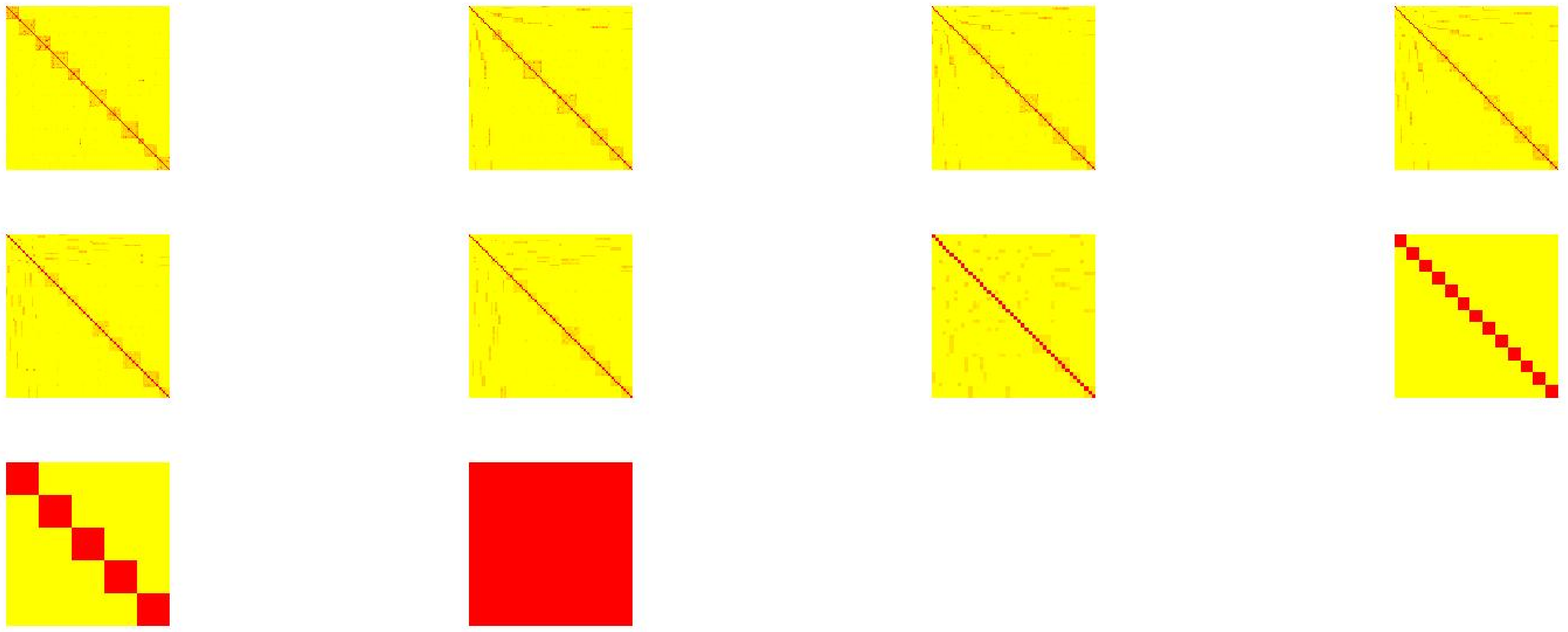}
		\caption{Affinity matrices created at different levels of hierarchy for $Net_{2}$ network. The number of block-diagonals in each subgraph represents  $k$ at that level of hierarchy.}
		\label{fig:subfig5}
	}
	\end{subfigure}
	\begin{subfigure}{\textwidth}
	{
		\centering
		\includegraphics[width=0.47\textwidth]{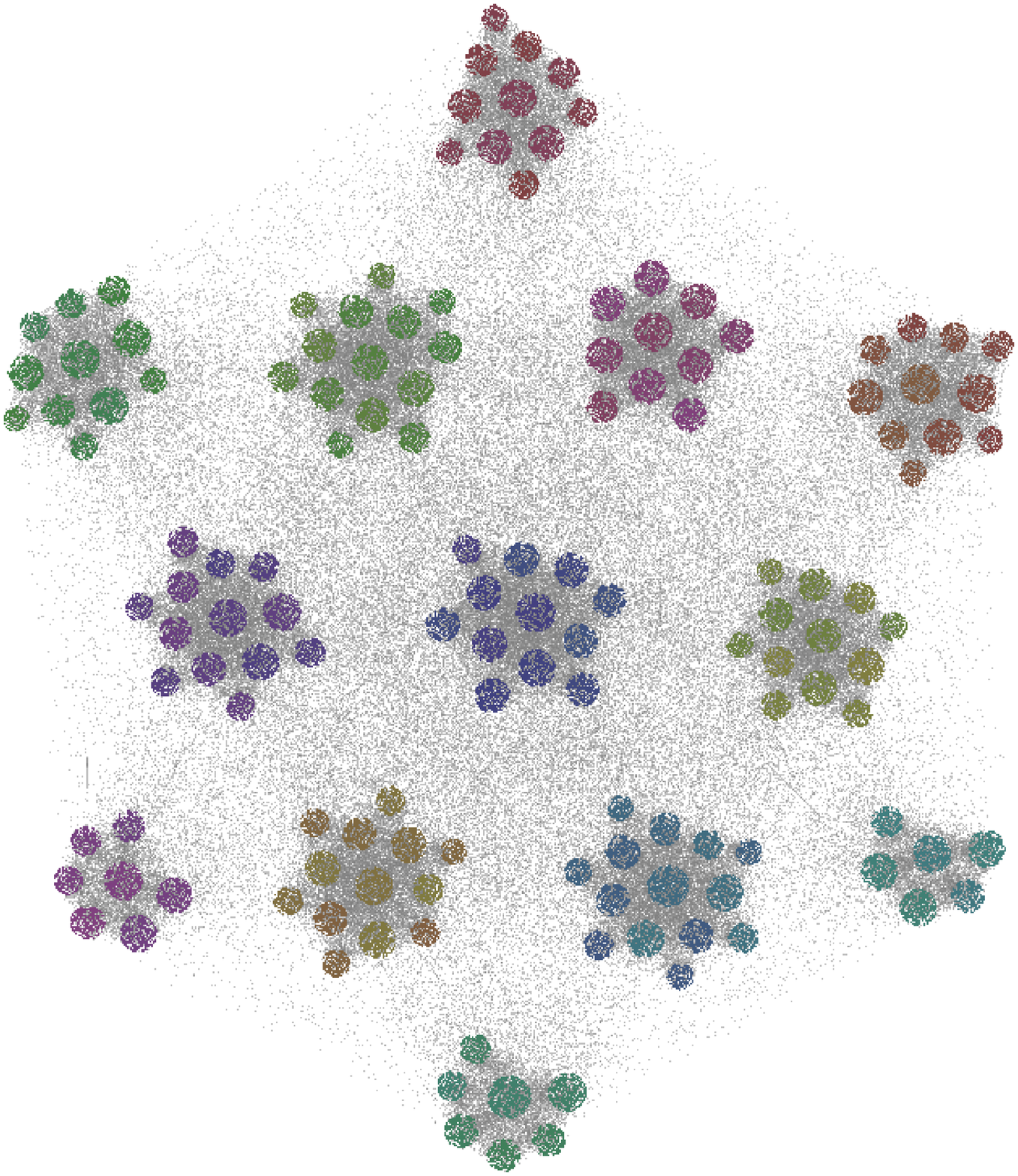}
		\includegraphics[width=0.47\textwidth]{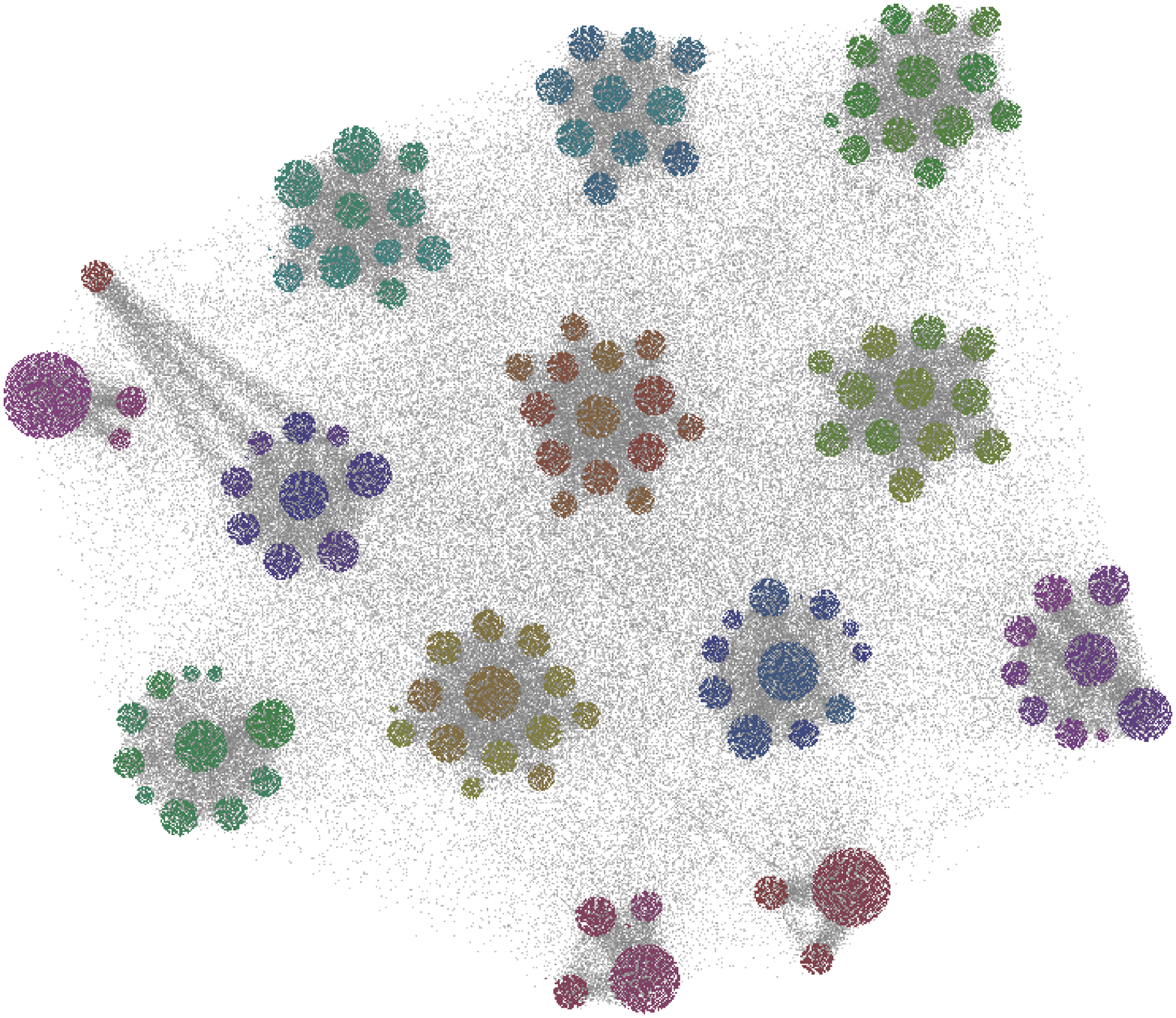}
		\caption{Original hierarchical network (left) and estimated hierarchical network (right) for synthetic network with $50,000$ nodes. The orientation and position of the communities might vary in the two plots. Original network has $3$ clusters with $11$ micro communities, $2$ clusters with $14$, $13$, $12$ and $7$ micro communities each, $1$ cluster with $10$ and another $1$ with $6$ micro communities. Estimated network has $3$ clusters with $11$ micro communities, $2$ clusters $13$, $10$ and $3$ micro communities each and $1$ cluster with $14$, $12$, $9$ and $4$ micro communities respectively.}
		\label{fig:subfig6}
	} 
	\end{subfigure}
	\caption{Result of MH-KSC algorithm on benchmark $Net_{2}$ network.}\label{fig:fig4}
\end{figure}
Figures \ref{fig:fig3} and \ref{fig:fig4} showcases the result of MH-KSC algorithm on the $Net_{1}$ and $Net_{2}$ respectively. From Figures \ref{fig:subfig3} and \ref{fig:subfig5}, we observe the affinity matrices generated corresponding to the test set for $Net_{1}$ and $Net_{2}$ respectively. From Figures \ref{fig:subfig4} and \ref{fig:subfig6}, we can observe the communities prevalent in the original network and the communities estimated by MH-KSC method for $Net_{1}$ and $Net_{2}$ respectively. In $Net_{1}$ there are $9$ macro communities and $37$ micro communities while in $Net_{2}$ there are $13$ macro communities and $141$ micro communities as depicted by Figures \ref{fig:subfig4} and \ref{fig:subfig6}.

Table \ref{table:t1} illustrates the first $10$ levels of hierarchy for $Net_{1}$ and $Net_{2}$ and evaluates the clusters obtained at each level of hierarchy w.r.t. quality metrics $ARI$, $VI$, $Q$ and $CC$. Higher values of $ARI$ (close to $1$) and lower values of $VI$ (close to $0$) represent good quality clusters. Both these external quality metrics are normalized as shown in \cite{rabbany}. Higher values of modularity ($Q$ close to $1$) and lower values of cut-conductance ($CC$ close to $0$) indicate better clustering information. 
\begin{table}[!ht]
\scriptsize
 \tabcolsep=0.05cm
  \centering
  \begin{tabular}{|c|c|c|c|c|c|c|c|c|c|c|}
  	\hline
  	 & \multicolumn{5}{|c|}{$Net_{1}$} & \multicolumn{5}{|c|}{$Net_{2}$} \\
    \hline
    \textbf{Hierarchy} & \textbf{$k$} & \textbf{$ARI$} & \textbf{$VI$} & \textbf{$Q$} & \textbf{$CC$} & \textbf{$k$} & \textbf{$ARI$} & \textbf{$VI$} & \textbf{$Q$} & \textbf{$CC$} \\
    \hline
    10 & - & - & - & - & - & \textbf{134} & \textbf{0.685} & \textbf{0.612} & 0.66 & \textbf{1.98e-05} \\
    9 & - & - & - & - & - & 112 & 0.625 & 0.643 & 0.685 & 1.99e-05 \\
    8 & - & - & - & - & - & 106 & 0.61 & 0.667 & 0.691 & 1.99e-05 \\
    7 & 63 & 0.972 & 0.11 & 0.62 & \textbf{4.74e-04} & 103 & 0.595 & 0.692 & 0.694 & \textbf{1.98e-05} \\
    6 & 40 & 0.996 & 0.018 & 0.668 & 4.86e-04 & 97 & 0.53 & 0.77 & 0.706 & 1.99e-05 \\
    5 & 39 & \textbf{0.996} & \textbf{0.016} & 0.669 & \textbf{4.834e-04} & 87 & 0.47 & 0.90 & 0.722 & 1.99e-05 \\
    4 & \textbf{37} & 0.965 & 0.056 & 0.675 & 4.856e-04 & 44 & 0.636 & 0.74 & \textbf{0.773} & 1.99e-05 \\
    3 & 15 & 0.878 & 0.324 & \textbf{0.765} & 5.021e-04 & \textbf{13} & \textbf{1.0} & \textbf{0.0} & \textbf{0.82} & 2.0e-05 \\
    2 & \textbf{9} & \textbf{1.0} & \textbf{0.0} & \textbf{0.786} & 5.01e-04 & 5 & 0.12 & 1.643 & 0.376 & 2.12e-05 \\
    1 & 1 & 0.0 & 2.19 & 0.0 & 5.0e-04 & 1 & 0.0 & 2.544 & 0.0 & 2.0e-05 \\
    \hline
  \end{tabular}
  \caption{Number of clusters ($k$) for top $10$ levels of hierarchy by MH-KSC method. The number of clusters close to the actual number, the best and second best results are highlighted. For $Net_{1}$ only $7$ levels of hierarchy are identified by MH-KSC, rest are represented by `-'. The MH-KSC method provides more insight by identifying several meaningful levels of hierarchy with good clusters w.r.t. quality metrics like $ARI$, $VI$, $Q$ and $CC$. }\label{table:t1}
\end{table}

Table \ref{table:t2} provides the result of Louvain, Infomap and OSLOM methods and compares it with the best levels of hierarchy for $Net_{1}$ and $Net_{2}$. The Louvain, Infomap and OSLOM methods require multiple runs as in each iteration they result in a different partition. We perform $10$ runs and report the mean results in Table \ref{table:t2}. From Table \ref{table:t2}, it can be observed that the best results for Louvain and Infomap methods generally occur at coarse levels of hierarchy w.r.t. to $ARI$, $VI$ and $Q$ metric. Thus, these two methods work well to identify macro communities. The Louvain method works the better than MH-KSC for $Net_{2}$ at macro and micro level. However, it cannot obtain similar quality micro communities when compared with MH-KSC method for $Net_{1}$ as inferred from Table \ref{table:t2}. The Infomap method performs the worst among all the methods w.r.t. detection of communities at finer levels of granularity. OSLOM performs well w.r.t. to locating both macro communities for $Net_{1}$ and micro communities for $Net_{2}$ as observed from Table \ref{table:t2}. It performs better than any method w.r.t. locating micro communities for $Net_{2}$ w.r.t. $ARI$ and $VI$ metric. However, it performs worst  while trying to identify the macro communities for the same benchmark network. The MH-KSC performs best on $Net_{1}$ while it performs better w.r.t. locating macro communities for $Net_{2}$. 
\begin{table}
\scriptsize
	\tabcolsep=0.05cm
  	\centering
  	\begin{tabular}{|c|c|c|c|c|c|c|c|c|c|c|c|c|}
  	\hline
  	\textbf{Method} & \multicolumn{6}{|c|}{$Net_{1}$} & \multicolumn{6}{|c|}{$Net_{2}$} \\
  	\hline
  	& \textbf{Level} & k & \textbf{$ARI$} & \textbf{$VI$}  & \textbf{$Q$} & \textbf{$CC$} & \textbf{Level} & k & \textbf{$ARI$} & \textbf{$VI$}  & \textbf{$Q$} & \textbf{$CC$} \\
  	\hline
  	\textbf{Louvain} & 2 & 32 & 0.84 & 0.215 & \textbf{0.693} & 4.87e-05 & 3 & 135 & 0.853 & 0.396 & \textbf{0.687} & \textbf{1.98e-05} \\
  	& 1 & 9 & \textbf{1.0} & \textbf{0.0} & \textbf{0.786} & \textbf{5.01e-04} & 1 & 13 & \textbf{1.0} & \textbf{0.0} & \textbf{0.82} & \textbf{2.0e-05} \\
  	\hline
  	 \textbf{Infomap} & 2 & 8 & 0.915 & 0.132 & 0.771 & 5.03e-04 & 3 & 590 & 0.003 & 8.58 & 0.003 & \textbf{1.98e-05} \\
  	 & 1 & 6 & 0.192 & 1.965 & 0.487 & 5.07e-04 & 1  & 13 & \textbf{1.0} & \textbf{0.0} & \textbf{0.82} & \textbf{2.0e-05}\\
  	\hline
  	\textbf{OSLOM} & 2 & 38 & 0.988 & 0.037 & 0.655 & 4.839e-04 & 2 & 141 & \textbf{0.96} & \textbf{0.214} & 0.64 & 2.07e-05 \\
  		  & 1 & 9 & \textbf{1.0} & \textbf{0.0} & \textbf{0.786} & \textbf{5.01e-04} & 1 & 29 & 0.74 & 0.633 & 076 & 2.08e-05 \\
  	\hline
  	\textbf{MH-KSC} & 5 & 39 & \textbf{0.996} & \textbf{0.016} & 0.67 & \textbf{4.83e-04} & 10 & 134 & 0.685 & 0.612 & 0.66 & \textbf{1.98e-05} \\
  	& 2 & 9 & \textbf{1.0} & \textbf{0.0} & \textbf{0.786} & \textbf{5.01e-04} & 3 & 13 & \textbf{1.0} & \textbf{0.0} & \textbf{0.82} & \textbf{2.0e-05} \\
  	\hline
  	\end{tabular}
  	\caption{$2$ best level of hierarchy obtained by Louvain, Infomap, OSLOM and MH-KSC methods on $Net_{1}$ and $Net_{2}$ benchmark networks. The best results w.r.t. various quality metrics when compared with the ground truth communities for each benchmark network is highlighted.}\label{table:t2}
\end{table}
%%%%%%%%%%%%%%%%%%%%%%%%%%%%%%%%%%%%%%%%%%%%%%%%
\subsection{Real-Life Network Experiments}
We experimented on $7$ real-life networks from the Stanford SNAP datasets \burl{http://snap.stanford.edu/data/index.html}. These networks are anonymous networks and are converted to undirected and unweighted networks before performing experiments on them. Table \ref{table:t3} provides information about topological characteristics of these real-life networks. The Fb and Epn networks are social networks, PGP is a trust based network, Cond is a collaboration network between researchers, Enr is an email network, Imdb is an actor-actor collaboration network and Utube is a web graph depicting friendship between the users of Youtube.
\begin{table}[!ht]
\scriptsize{
	\centering
	\hfill{}
	\begin{tabular}{|c|c|c|c|}
	\hline
	\textbf{Network} & \textbf{Nodes} & \textbf{Edges} & \textbf{CCF}\\
	\hline
	\textbf{Facebook (Fb)} & 4,039 & 88,234 & 0.6055 \\
	\textbf{PGPnet (PGP)} & 10,876 & 39,994 & 0.008 \\
    \textbf{Cond-mat (Cond)} & 23,133 & 186,936 & 0.6334 \\
	\textbf{Enron (Enr)} & 36,692 & 367,662 & 0.497 \\
	\textbf{Epinions (Epn)} & 75,879 & 508,837  & 0.1378 \\
	\textbf{Imdb-Actor (Imdb)} & 383,640 & 1,342,595 & 0.453 \\
	\textbf{Youtube (Utube)} & 1,134,890 & 2,987,624 & 0.081\\
	\hline
	\end{tabular}
	}
	\hfill{}
	\caption{Nodes (V), Edges (E) and Clustering Coefficients (CCF) for each network}\label{table:t3}
\end{table}

In case of real-life networks the true hierarchical structure is not known beforehand. Hence, it is important to show whether they exhibit hierarchical organization which can be tested by identifying good quality clusters w.r.t. internal quality metrics like $Q$ and $CC$ at multiple levels of hierarchy.

We showcase the results for $10$ levels of hierarchy in a bottom-up fashion for the MH-KSC method in Table \ref{table:t4}. The coarsest level of hierarchy has all nodes in one community and is not very insightful. Clusters at very coarse levels of granularity comprises giant connected components. So, it is more meaningful to give more emphasis to fine grained clusters at lower levels of hierarchy. To show that real-life networks exhibit hierarchy we evaluate our proposed MH-KSC approach in Table \ref{table:t4}.
\begin{table}
\scriptsize
	\tabcolsep=0.005cm
	\centering
	\begin{tabular}{|c|c|c|c|c|c|c|c|c|c|c|c|c|}
	\hline
	& & \multicolumn{10}{|c|}{Hierarchical Organization} \\
	\hline
	 \textbf{Network} & \textbf{Metrics} & \textbf{Level 12} & \textbf{Level 11} & \textbf{Level 10} & \textbf{Level 9} & \textbf{Level 8} & \textbf{Level 7} & \textbf{Level 6} & \textbf{Level 5} & \textbf{Level 4} & \textbf{Level 3} \\
	\hline 
	 & $k$ & 358 & 192 & 152 & 121 & 105 & 90 & 71 & 43 & 37 & 21 \\
	 Fb & $Q$ & 0.604 & 0.764 & 0.769 & 0.789 & 0.792 & 0.81 & 0.812 & 0.818 & 0.821 & \textbf{0.83} \\
	 & $CC$ & \textbf{2.47e-05} & 1.56e-04 & 2.38e-04 & 1.91e-04 & 1.95e-04 & 1.63e-04 & 2.16e-04 & 1.76e-04 & 2.44e-04 & 2.4e-04 \\
	 \hline
	 & $k$ & 345 & 274 & 202 & 156 & 129 & 83 & 59 & 46 & 24 & 19 \\
	 PGP & $Q$ & 0.682 & 0.693 & 0.705 & 0.715 & 0.725 & 0.727 & 0.728 & \textbf{0.729} & 0.701 & 0.698 \\
	 & $CC$ & 8.48e-05 & 9.84e-05 & 5.88e-05 & 1.38e-04 & 7.2e-05 & 8.03e-05 & 1.0e-04 & 1.07e-04 & 4.13e-04 & \textbf{4.89e-05}\\
	 \hline 
	 & $k$ & 2676 & 1171 & 621 & 324 & 171 & 102 & 80 & 58 & 41 & 24 \\
	 Cond & $Q$ & 0.5 & 0.567 & 0.586 & 0.611 & \textbf{0.615} & 0.614 & 0.582 & 0.582 & 0.574 & 0.515 \\
	 & $CC$ & 2.49e-05 & 2.6e-05 & 3.7e-05 & 3.52e-05 & 3.6e-05 & 5.86e-05 & 2.37e-05 & 3.45e-05 & 1.43e-05 & \textbf{1.4e-05} \\
	 \hline
	 & $k$ & 2208 & 1002 & 464 & 303 & 211 & 163 & 119 & 76 & 59 & 48 \\
	 Enr & $Q$ & 0.30 & 0.388 & 0.444 & 0.451 & \textbf{0.454} & 0.427 & 0.43 & 0.325 & 0.328 & 0.271 \\
	 & $CC$ & \textbf{1.19e-05} & 3.18e-05 & 3.1e-05 & 5.3e-05 & 7.04e-05 & 2.69e-04 & 2.2e-03 & 1.651e-04 & 2.56e-05 & 5.46e-05\\
	 \hline
	 & $k$ & 8808 & 3133 & 1964 & 957 & 351 & 220 & 166 & 97 & 66 & 26 \\
	 Epn & $Q$ & 0.105 & 0.156 & 0.158 & 0.176 & 0.184 & 0.183 & \textbf{0.186} & 0.184 & 0.146 & 0.006 \\
	 & $CC$ & \textbf{1.4e-06} & 3.1e-06 & 6.4e-06 & 7.0e-06 & 9.5e-06 & 1.26e-05 & 7.0e-06 & 9.0e-06 & 2.42e-05 & 7.8e-06 \\
	 \hline 
	 & $k$ & 7431 & 1609 & 890 & 468 & 313 & 200 & 130 & 72 & 46 & 21 \\
	 Imdb & $Q$ & 0.357 & 0.47 & 0.473 & 0.485 & 0.503 & \textbf{0.521} & 0.508 & 0.514 & 0.513 & 0.406 \\
	 & $CC$ & 1.43e-06 & 2.78e-06 & 2.79e-06 & 5.6e-06 & 4.24e-06 & 5.6e-06 & 6.42e-06 & 1.99e-06 & 7.46e-06 & \textbf{9.2e-07}\\
	 \hline
	 & $k$ & 9984 & 2185 & 529 & 274 & 180 & 131 & 100 & 71 & 46 & 26 \\
	 Utube & $Q$ & 0.524 & 0.439 & 0.679 & \textbf{0.682} & 0.599 & 0.491 & 0.486 & 0.483 & 0.306 & 0.303 \\
	 & $CC$ & \textbf{2.65e-07} & 3.0e-07 & 1.3e-06 & 2.4e-06 & 1.0e-06 & 7.6e-06 & 1.03e-5 & 1.07e-05 & 2.33e-05 & 1.55e-04 \\  
	 \hline 
	\end{tabular}
	\caption{Results on MH-KSC algorithm on $7$ real-life networks using quality metrics $Q$ and $CC$. The best results corresponding to each metric for individual networks are highlighted.}\label{table:t4}	
\end{table}
%%%%%%%%%%%%%%%%%%%%%%%%%%%%%%%%%%%%%%%%%%%%%%%%%%%%%%%%%%%%%%%%%%%%%
\begin{table}
\scriptsize
	\tabcolsep=0.01cm
	\centering
	\begin{tabular}{|c|c|c|c|c|c|c|c|}
	\hline
	& & \multicolumn{6}{|c|}{Hierarchical Organization} \\
	\hline 
	\textbf{Network} & \textbf{Metrics} & \textbf{Level 6} & \textbf{Level 5} & \textbf{Level 4} & \textbf{Level 3} & \textbf{Level 2} & \textbf{Level 1} \\
	\hline
	& $k$ & - & - & - & 225 & 155 & 151 \\
	Fb & $Q$ & - & - & - & 0.82 & 0.846 & \textbf{0.847} \\
	& $CC$ & - & - & - & \textbf{9.88e-05} & 1.33e-04 & 1.32e-04 \\
	\hline
	& $k$ & - & - & 2392 & 566 & 154 & 100 \\
	PGP & $Q$ & - & - & 0.705 & 0.857 & 0.882 & \textbf{0.884} \\
	& $CC$ & - & - & \textbf{4.95e-05} & 8.66e-05 & 6.8e-05 & 1.0e-04 \\
	\hline
	& $k$ & - & - & 6732 & 1825 & 1066 & 1011 \\
	Cond & $Q$ & - & - & 0.56 & 0.7 & 0.731 & \textbf{0.732} \\
	& $CC$ & - & - & \textbf{1.56e-05} & 2.97e-05 & 3.49e-05 & 4.15e-05 \\
	\hline 
	& $k$ & - & - & 4001 & 1433 & 1237 & 1230 \\
	Enr & $Q$ & - & - & 0.546 & 0.608 & 0.613 & \textbf{0.614} \\
	& $CC$ & - & - & \textbf{1.28e-05} & 1.88e-05 & 4.58e-05 & 6.48e-05 \\
	\hline
	& $k$ & 10351 & 2818 & 1574 & 1325 & 1301 & 1300 \\
	Epn & $Q$ & 0.287 & 0.319 & 0.323 & 0.324 & 0.324 & \textbf{0.324} \\
	& $CC$ & \textbf{1.86e-06} & 4.2e-06 & 4.25e-06 & 5.57e-06 & 6.75e-06 & 1.13e-05 \\
	\hline
	& $k$ & - & 22613 & 4544 & 3910 & 3815 & 3804 \\
	Imdb & $Q$ & - & 0.591 & 0.727 & 0.729 & 0.729 & \textbf{0.729} \\ 
	& $CC$ & - & \textbf{1.0e-06} & 1.0e-06 & 1.85e-06 & 2.5e-06 & 2.82e-06 \\
	\hline
	& $k$ & 33623 & 11587 & 6964 & 6450 & 6369 & 6364 \\
	Utube & $Q$ & 0.696 & 0.711 & 0.714 & 0.715 & 0.715 & \textbf{0.715} \\
	& $CC$ & \textbf{1.38e-06} & 2.22e-06 & 3.25e-06 & 3.98e-06 & 4.06e-06 & 9.96e-06 \\
	\hline
	\end{tabular}
	\caption{Results of Louvain method on $7$ real-life networks indicating the top $6$ levels of hierarchy. The best results are highlighted and `-' is used in case the metric is not applicable due to absence of partitions.}\label{table:t5}
\end{table}
\begin{table}
\scriptsize
	\tabcolsep=0.01cm
	\centering
	\begin{tabular}{|c|c|c|c|c|c|c|c|c|}
	\hline
	& & \multicolumn{2}{|c|}{Infomap} & \multicolumn{5}{|c|}{OSLOM} \\
	\hline
	& & \multicolumn{2}{|c|}{Hierarchical Info} & \multicolumn{5}{|c|}{Hierarchical Info} \\
	\hline
	\textbf{Network} & \textbf{Metrics} & \textbf{Level 2} & \textbf{Level 1} & \textbf{Level 5} & \textbf{Level 4} & \textbf{Level 3} & \textbf{Level 2} & \textbf{Level 1} \\
	\hline
	& $k$ & 325 & 131 & - & 161 & 50 & 27 & 21 \\ 
	Fb & $Q$ & 0.055 & \textbf{0.763} & - & 0.045 & 0.133 & 0.352 & \textbf{0.415} \\
	& $CC$ & \textbf{2.86e-05} & 2.3e-04 & - & \textbf{2.0e-04} & 2.0e-04 & 3.0e-04 & 3.0e-04\\
	\hline
	& $k$ & 85 & 65 & 431 & 143 & 51 & 48 & 45 \\
	PGP & $Q$ & 0.041 & \textbf{0.862} & 0.748 & \textbf{0.799} & 0.709 & 0.709 & 0.709 \\
	& $CC$ & 1.66e-04 & \textbf{1.40e-04} & 1.74e-04 & \textbf{5.32e-05} & 2.06e-04 & 1.56e-04 & 6.64e-05 \\
	\hline
	& $k$ & 1009 & 173 & 4092 & 2211 & 1745 & 1613 & 1468 \\
	Cond & $Q$ & \textbf{0.648} & 0.027 & 0.483 & 0.574 & 0.615 & \textbf{0.615} & 0.05 \\
	& $CC$ & \textbf{1.71e-05} & 2.78e-05 &	1.77e-05 & 2.48e-05 & 3.04e-05 & 6.56e-05 & \textbf{1.16e-05} \\
	\hline
	& $k$ & 1920 & 1084 & - & 3149 & 2177 & 2014 & 1970 \\
	Enr & $Q$ & 0.015 & \textbf{0.151} & - & 0.317 & 0.382 & 0.412 & \textbf{0.442} \\
	& $CC$ & \textbf{1.83e-05} & 8.39e-04 & - & \textbf{1.75e-05} & 4.96e-05 & 9.92e-05 & 7.22e-05 \\
	\hline
	& $k$ & 14170 & 50 & 1693 & 584 & 206 & 30 & 25 \\
	Epn & $Q$ & 5.3e-06 & \textbf{4.48e-04} & 0.162 & 0.226 & \textbf{0.239} & 0.098 & 0.019 \\
	& $CC$ & \textbf{3.97e-06} & 4.63e-05 & 1.23e-05 & 9.75e-06 & 2.45e-05 & 8.2e-06 & \textbf{7.9e-06} \\
	\hline
	& $k$ & 14308 & 3238 & - & 7469 & 2639 & 2017 & 2082 \\
	Imdb & $Q$ & 0.04 & \textbf{0.707} & - & 0.045 & 0.092 & 0.1 & 0.115 \\
	& $CC$ & \textbf{1.23e-06} & 4.72e-06 & - & \textbf{1.35e-06} & 2.03e-06 & 7.95r-06 & 1.17e-05 \\
	\hline
	& $k$ & 10703 & 976 & 18539 & 6547 & 4184 & 2003 & 1908\\
	Utube & $Q$ &  0.035 & \textbf{0.698} & 0.396 & 0.53 & \textbf{0.588} & 0.487 & 0.027 \\
	& $CC$ & \textbf{1.38e-06} & 5.56e-06 & 1.52e-06 & 3.1e-07 & \textbf{2.72e-07} & 6.1e-06 & 5.69e-06 \\
	\hline
	\end{tabular}
	\caption{Results of Infomap and OSLOM methods. The best results for each method corresponding to each network is highlighted and `-' represent not applicable cases.}\label{table:t6}
\end{table}

We compare MH-KSC algorithm with Louvain \cite{blondel}, Infomap \cite{rosvall:bergstrom} and OSLOM \cite{lanchichinetti:radicchi}. We perform $10$ runs for each of these methods as they generate a separate partition each time when they are executed. The mean results of Louvain method is reported in Table \ref{table:t5}. Table \ref{table:t6} showcases the results for Infomap and OSLOM method.

From Table \ref{table:t5} it is evident that the Louvain method works best w.r.t. the modularity ($Q$) criterion. This aligns with methodology as it is trying to optimize for $Q$. However, the Louvain method always performs worse than MH-KSC algorithm w.r.t. cut-conductance $CC$ as observed from Tables \ref{table:t4} and \ref{table:t5}. Another issue with the Louvain method is that except for the Fb and PGP networks it is not able to detect ($<1000$ clusters) high quality clusters at coarser levels of granularity. This is attributed to the resolution limit problem suffered by Louvain method. From Table \ref{table:t6} we observe that the Infomap method produces only $2$ levels of hierarchy. In most of the cases, the clusters at one level of hierarchy perform good w.r.t. only $1$ quality metric except the PGP and Cond networks. The difference between the quality of the clusters at the $2$ levels of hierarchy is quite drastic. This reflects that the Infomap method is not very consistent w.r.t. various quality metrics. 

We compare the performance of MH-KSC method with OSLOM in detail. From Tables \ref{table:t4} and \ref{table:t5} we observe that the MH-KSC technique outperforms OSLOM w.r.t. both quality metrics for Fb, Enr, Imdb and Utube networks while OSLOM does the same only for Cond network. In case of PGP, Cond and Epn networks OSLOM results in better $Q$ than MH-KSC. However, MH-KSC approach has better $CC$ value for PGP and Epn networks. For large scale networks like Enr, Imdb and Utube, OSLOM cannot identify good quality coarser clusters i.e. number of clusters detected are always $>1000$.
\subsection{Visualization and Illustrations}

\begin{figure}[!ht]
	\begin{subfigure}{\textwidth}
	{
		\centering
		\includegraphics[width=1.025\textwidth]{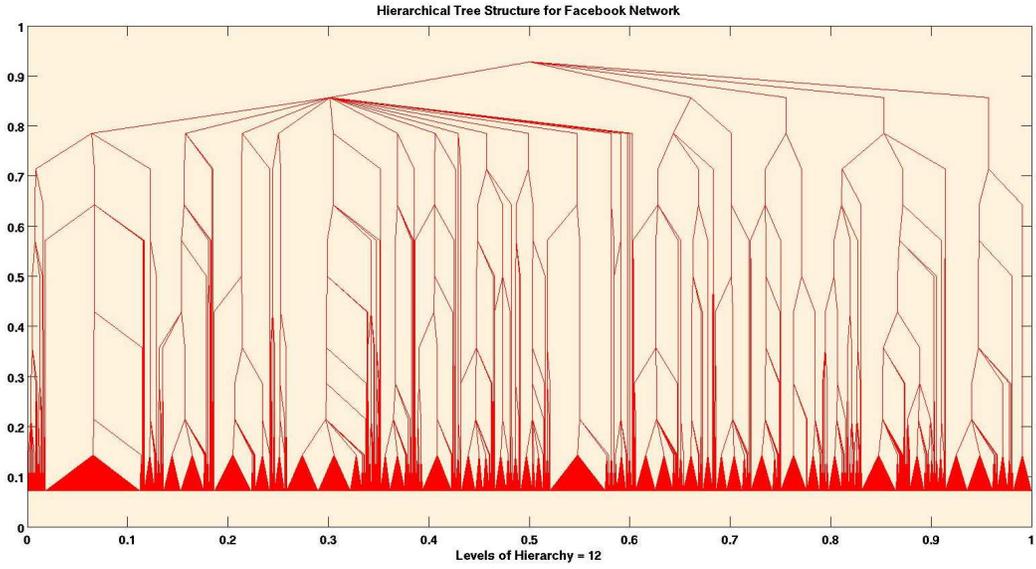}
		\caption{Multilevel Hierarchical Organization for Fb network}
		\label{fig:subfig7}
	}
	\end{subfigure}
	\begin{subfigure}{\textwidth}
	{
		\centering
		\includegraphics[width=1.025\textwidth]{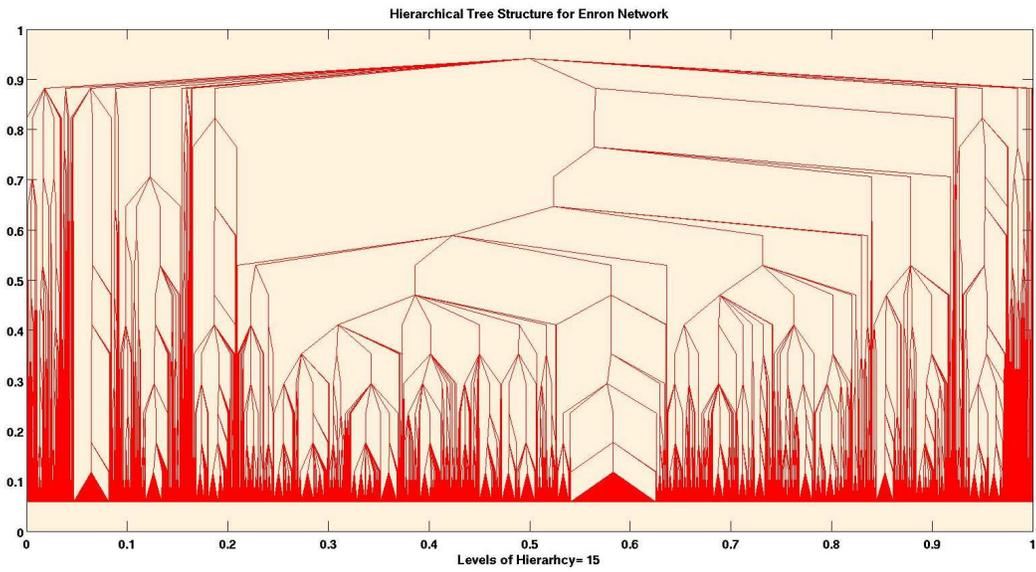}
		\caption{Multilevel Hierarchical Organization for Enr network}
		\label{fig:subfig8}
	}
	\end{subfigure}
	\caption{Tree based visualization of the multilevel hierarchical organization prevalent in $2$ real-life networks .}\label{fig:fig5}
\end{figure}

We provide a tree based visualization of the multilevel hierarchical organization for Fb and Enr networks in Figure \ref{fig:fig5}. The hierarchial structure is depicted as tree for Fb and Enr network in Figures \ref{fig:subfig7} and \ref{fig:subfig8} respectively.

\begin{figure}[!ht]
		\begin{subfigure}{0.48\textwidth}
		{	
			\centering
			\includegraphics[width=\textwidth]{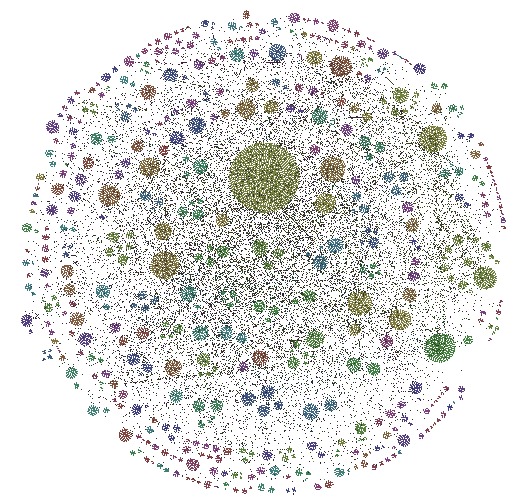}
			\caption{Many Micro Communities at Finer Levels}
			\label{fig:subfig9}
		}
		\end{subfigure}
			\begin{subfigure}{0.48\textwidth}
		{	
			\centering
			\includegraphics[width=\textwidth]{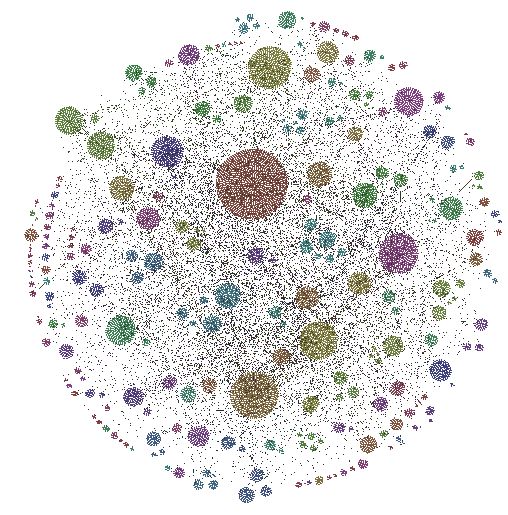}
			\caption{Less Micro Communites at Intermediate Levels}
			\label{fig:subfig10}
		}
		\end{subfigure}
		\begin{subfigure}{0.48\textwidth}
		{	
			\centering
			\includegraphics[width=\textwidth]{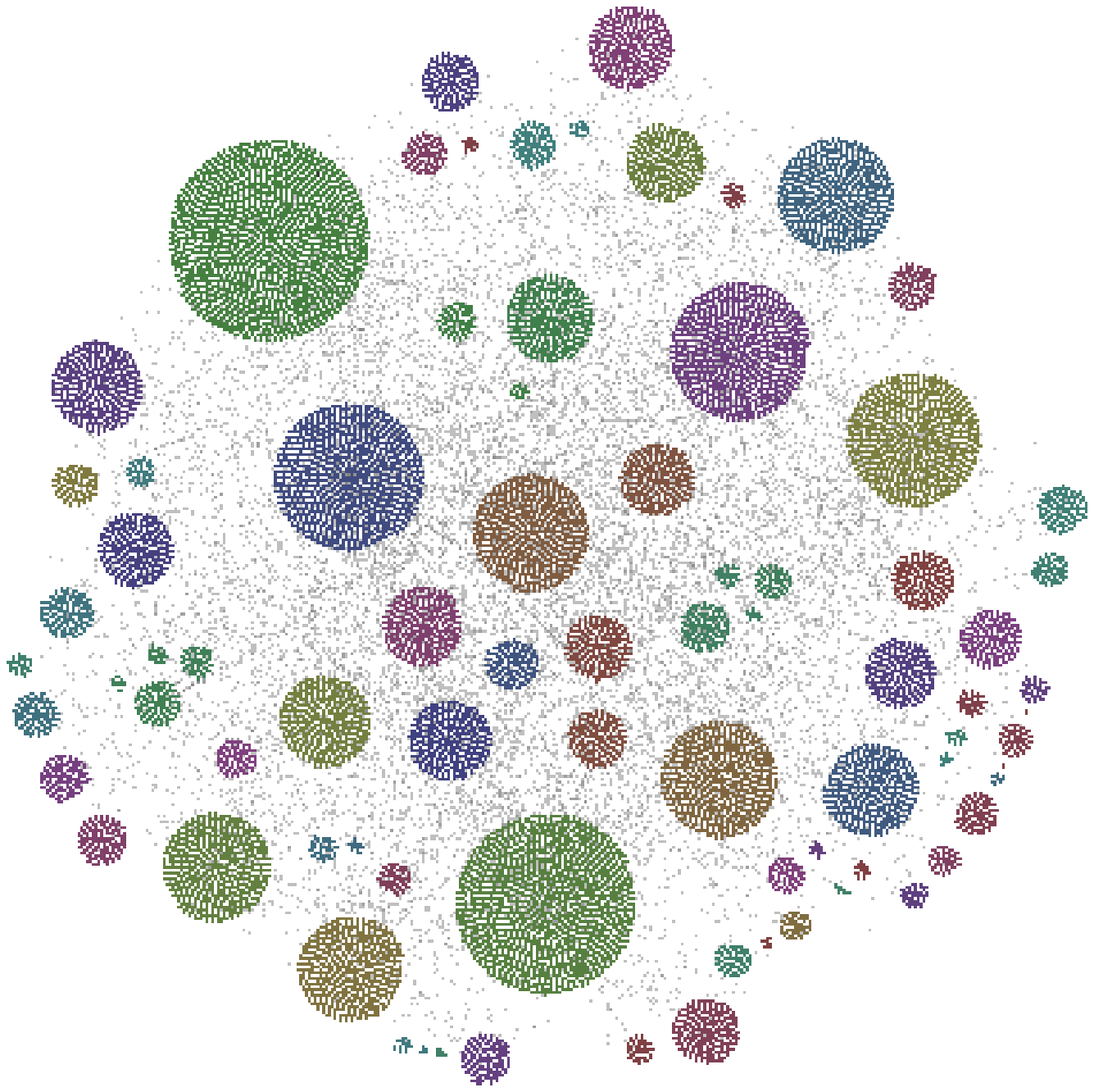}
			\caption{Few Micro and Few Macro Communities at Intermediate Levels}
			\label{fig:subfig11}
		}
		\end{subfigure}
		\begin{subfigure}{0.48\textwidth}
		{	
			\centering
			\includegraphics[width=\textwidth]{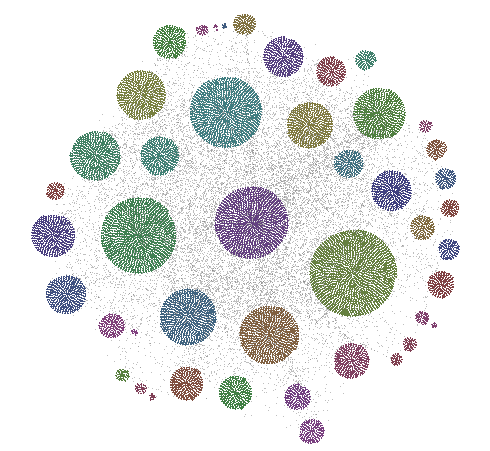}
			\caption{Some Pre-dominant Macro Communities at Coarser Levels}
			\label{fig:subfig12}
		}
		\end{subfigure}
		\caption{Results of the \textbf{MH-KSC} algorithm for the PGP network. Clusters with same colour are part of one community.}
		\label{fig:fig6}
	\end{figure}

\begin{figure}
		\begin{subfigure}{0.32\textwidth}
		{
			\centering
			\includegraphics[width=\textwidth]{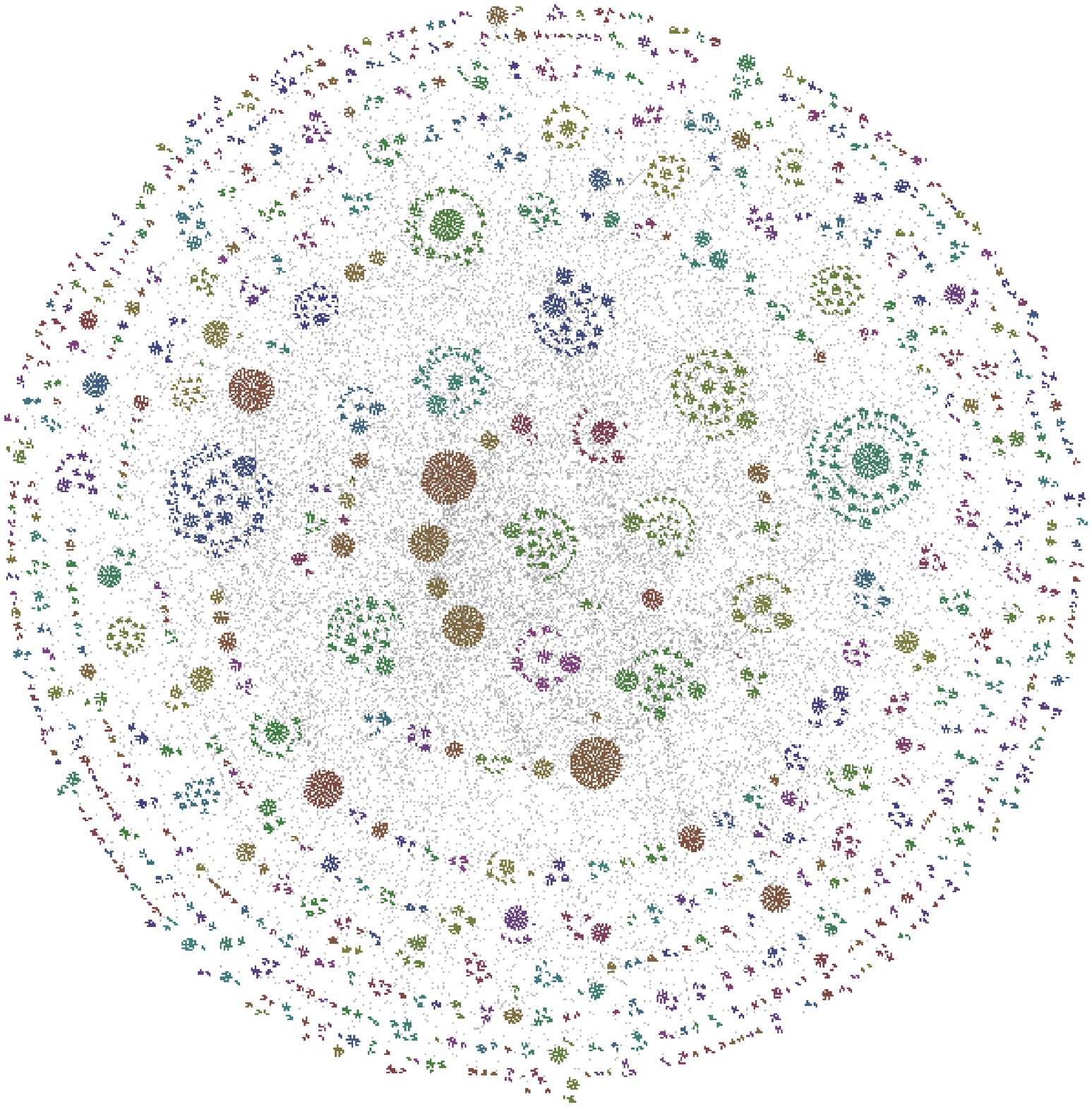}
			\caption{Many Micro Communities at Finer Levels for \textbf{Louvain} Method}
			\label{fig:subfig12}
		}
		\end{subfigure}
		\begin{subfigure}{0.32\textwidth}
		{
			\centering
			\includegraphics[width=\textwidth]{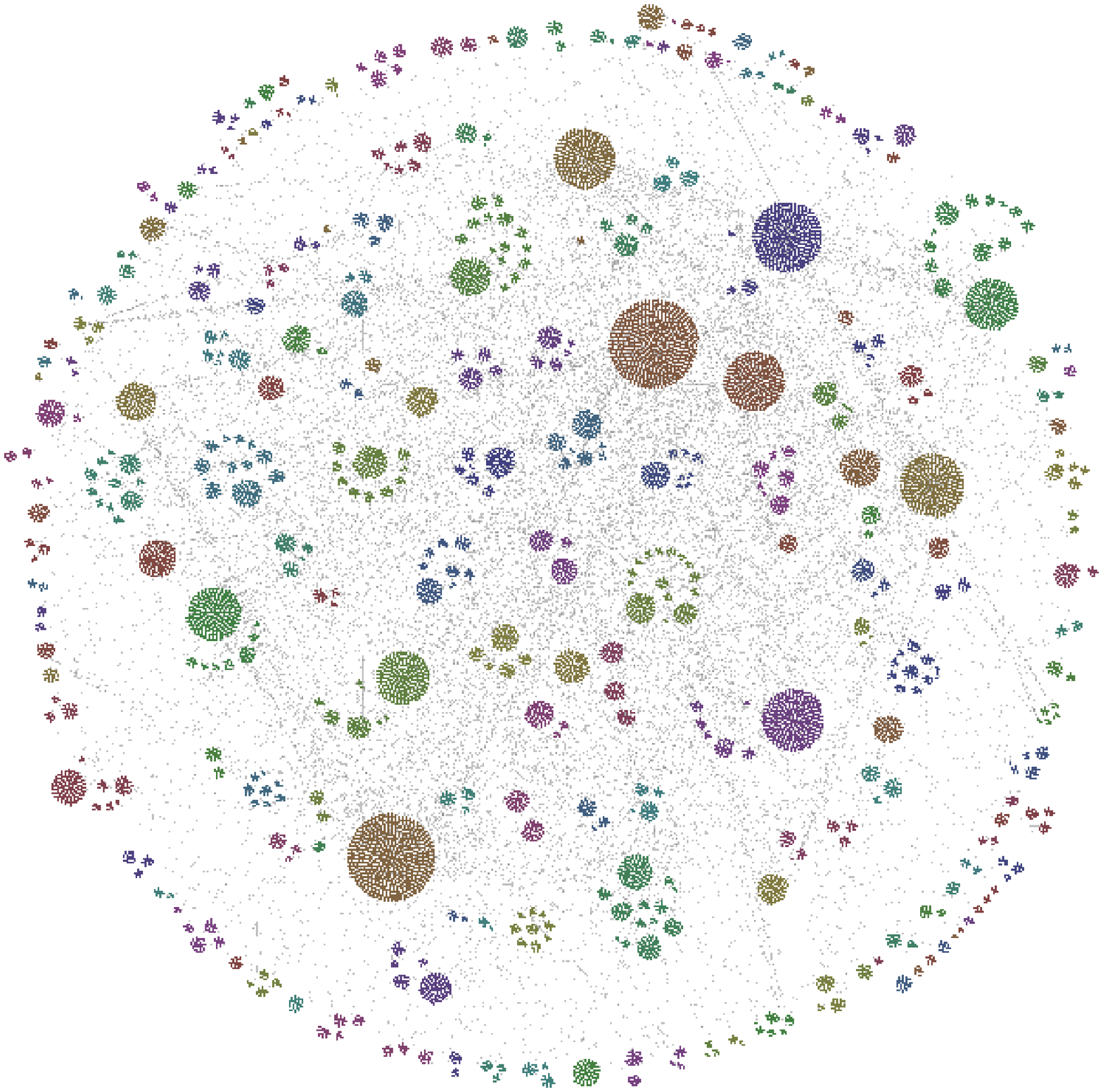}
			\caption{Few Micro Communities at Intermediate Levels for \textbf{Louvain} Method}
			\label{fig:subfig13}
		}
		\end{subfigure}
		\begin{subfigure}{0.32\textwidth}
		{
			\centering
			\includegraphics[width=\textwidth]{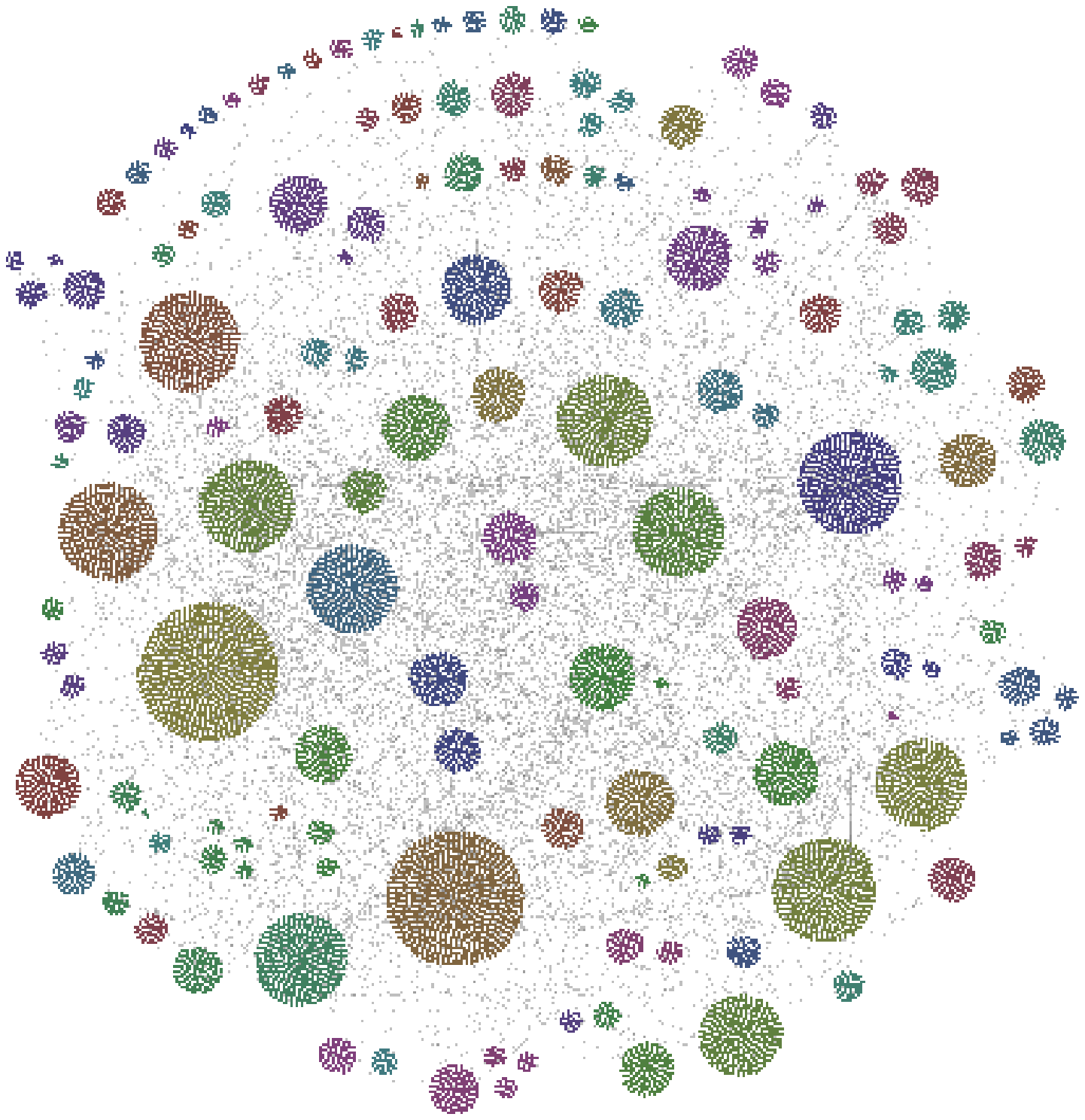}
			\caption{Few Micro and Macro Communities at Coarser Levels for \textbf{Louvain} Method}
			\label{fig:subfig14}
		}
		\end{subfigure}
		\begin{subfigure}{\textwidth}
		{
			\centering
			\includegraphics[width=0.32\textwidth]{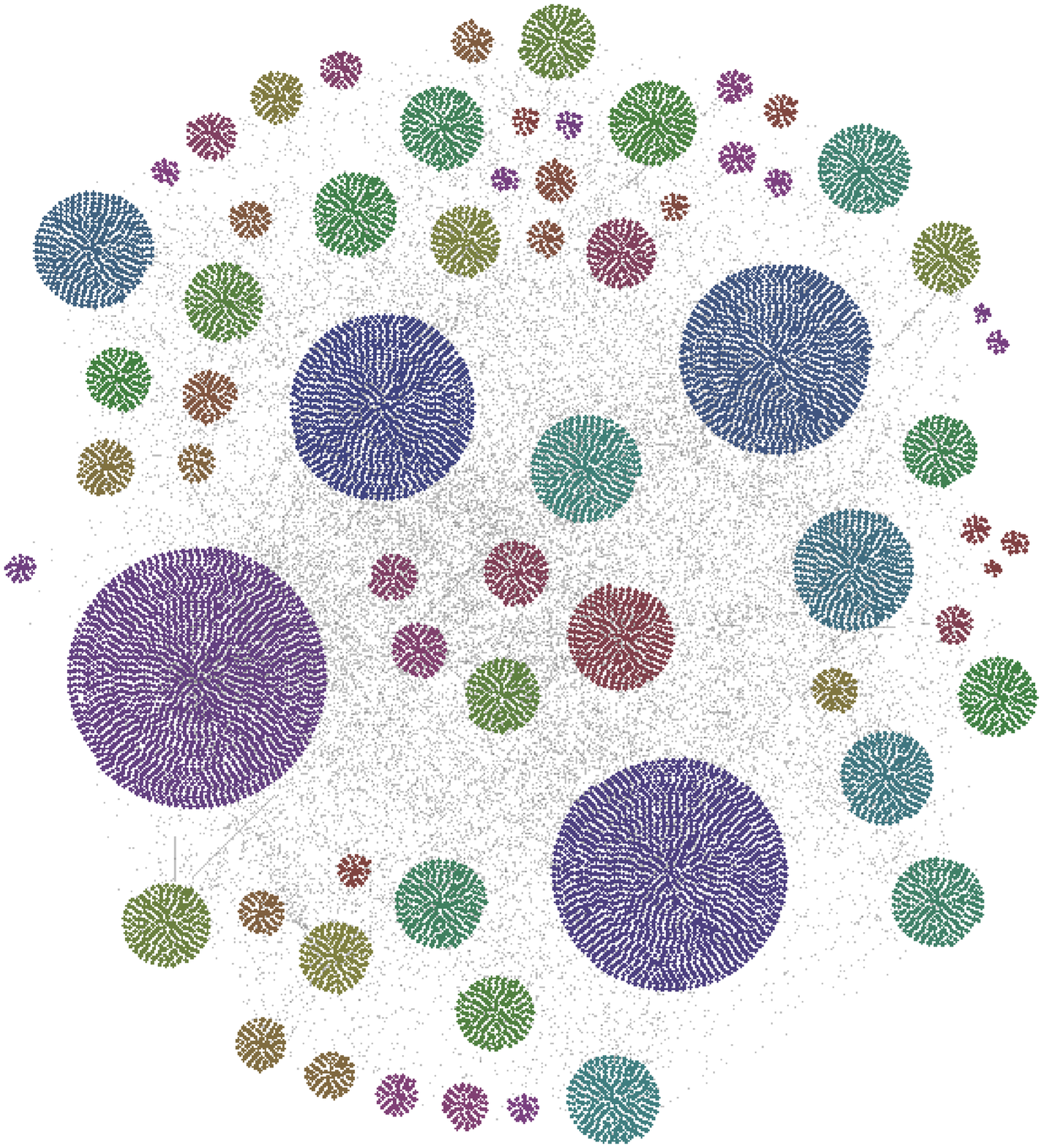}
			\caption{Some Macro Communities at Coarser Levels for \textbf{Infomap} Method}
			\label{fig:subfig15}
		}
		\end{subfigure}
		\begin{subfigure}{0.5\textwidth}
		{
			\centering
			\includegraphics[width=0.64\textwidth]{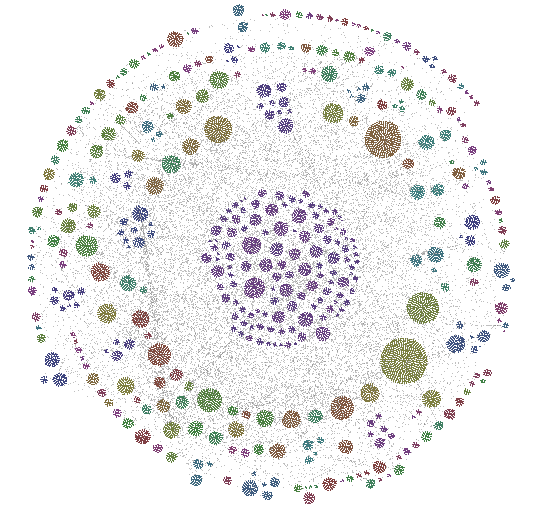}
			\caption{Many Micro Communities at Finer Levels for \textbf{OSLOM} Method}
			\label{fig:subfig16}
		}
		\end{subfigure}
		\begin{subfigure}{0.5\textwidth}
		{
			\centering
			\includegraphics[width=0.64\textwidth]{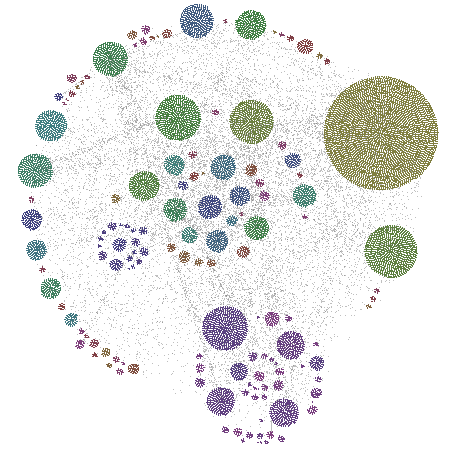}
			\caption{Few Micro and Macro Communities at Coarser Levels for \textbf{OSLOM} Method}
			\label{fig:subfig17}
		}
		\end{subfigure}
		\caption{Results of Louvain, Infomap and OSLOM methods for PGP network. Clusters with same colour are part of one community. Louvain method can only provide results with few macro and many micro communities. Infomap method only has $2$ coarse levels of hierarchy. OSLOM can provide both micro and macro communities but cannot detect as many meaningful intermediate layers as the MH-KSC method.}
		\label{fig:fig7}.
	\end{figure}

We plot the results corresponding to fine, intermediate and coarse levels of hierarchy for PGP network using the software provided in \cite{lanchichinetti:radicchi}. The software requires all the nodes in the network along with $2$ levels of hierarchy. In Figure \ref{fig:fig6} we plot the results for PGP net corresponding to MH-KSC algorithm using $2$ fine, $4$ intermediate and $2$ coarse levels of the hierarchical organization. For Louvain method we use $4^{th}$ and $3^{rd}$ level of hierarchy as inputs for the finest level, $3^{rd}$ and $2^{nd}$ level of hierarchy as inputs for intermediate level and $2^{nd}$ and $1^{st}$ level of hierarchy as inputs for coarsest level plot. The Infomap method only generates $2$ level of hierarchy which correspond to a coarse level plot. Similarly, for OSLOM we plot a fine and coarse level plot. The results for Louvain, Infomap and OSLOM methods are depicted in Figure \ref{fig:fig7}. 
	%%%%%%%%%%%%%%%%%%%%%%%%%%%%%%%%%%%%%%%%%%%%%%%%%%%%%%%%%%
%%%%%%%%%%%%%%%%%%%%%%%%%%%%%%%%%%%%

Figures \ref{fig:fig6} and \ref{fig:fig7} shows that MH-KSC algorithm allows to depict richer structures than the other methods. It has more flexibility and allows the visualization at coarser, intermediate and finer levels of granularity. From Figures \ref{fig:subfig12},\ref{fig:subfig13}, \ref{fig:subfig14} and Table \ref{table:t5}, we observe that the Louvain method can only detect quality clusters at finer levels of granularity and cannot detect less than $1,00$ communities. While the Infomap method can only locate giant connected components for the PGP network as observed from Figure \ref{fig:subfig15} and Table \ref{table:t6}. The OSLOM method also seems to work reasonably well as observed from Figures \ref{fig:subfig16} and \ref{fig:subfig17}. However, it detects fewer levels of hierarchy and thus has less flexibility in terms of selection for the level of hierarchy than the proposed MH-KSC approach. 

\begin{figure}[!ht]
	\begin{subfigure}{0.5\textwidth}
	{
		\centering
		\includegraphics[width=\textwidth]{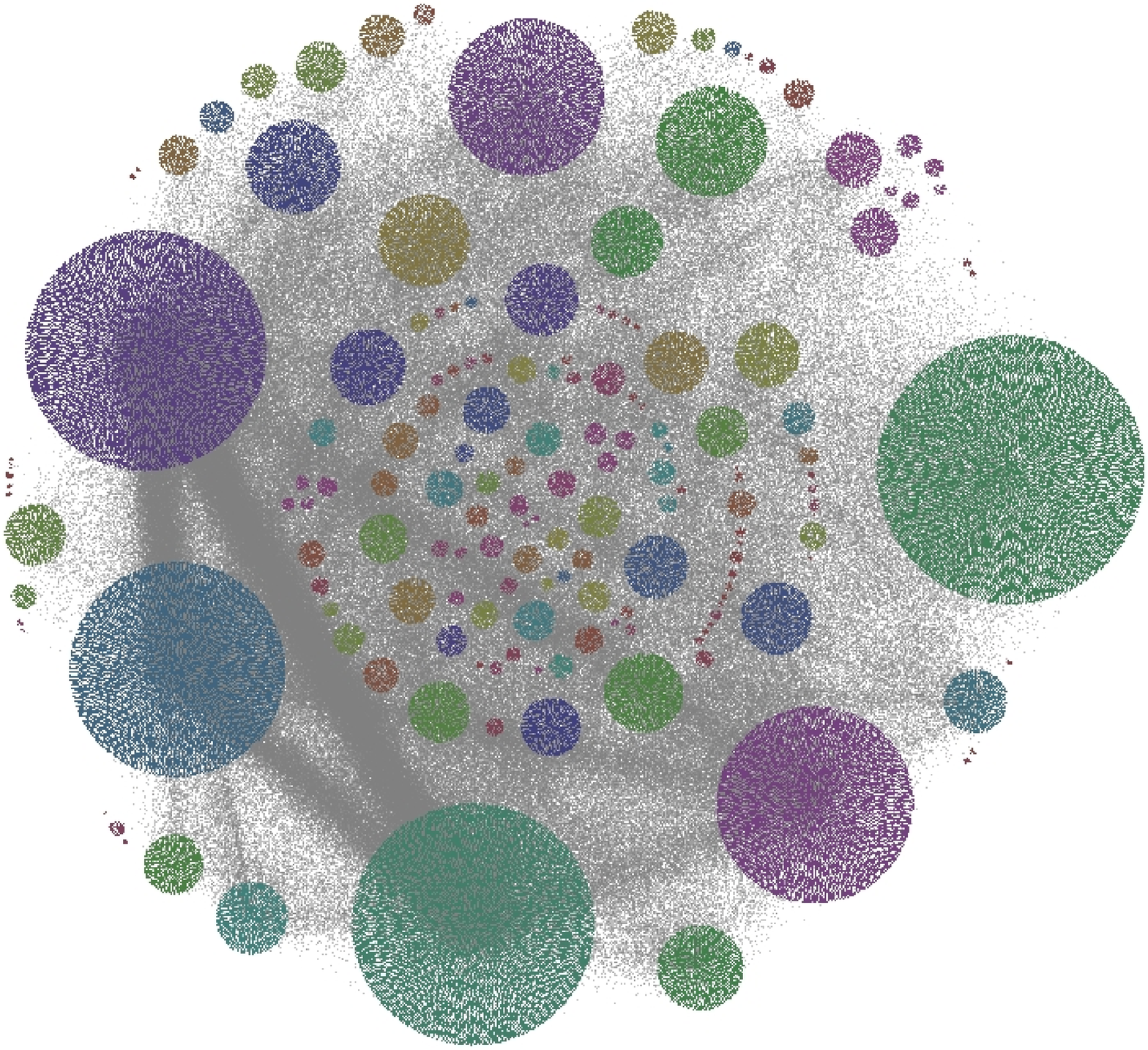}
		\caption{Best Result for MH-KSC algorithm}
		\label{fig:subfig18}
	}
	\end{subfigure}
		\begin{subfigure}{0.5\textwidth}
	{
		\centering
		\includegraphics[width=\textwidth]{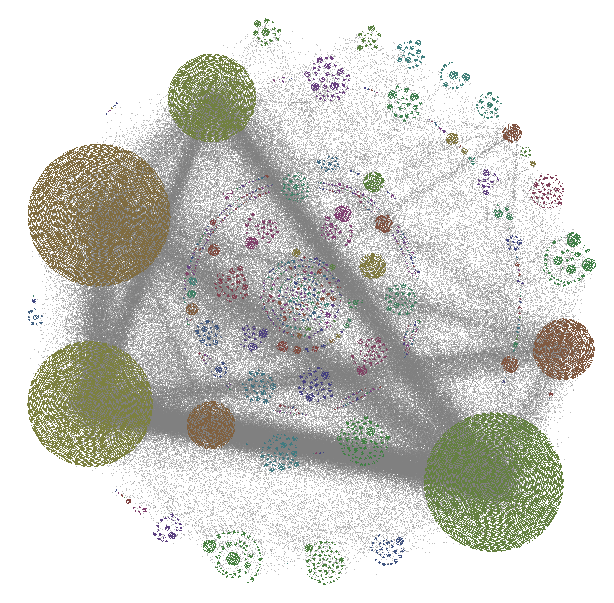}
		\caption{Best Result for Louvain method}
		\label{fig:subfig19}
	}
	\end{subfigure}
	\begin{subfigure}{0.5\textwidth}
	{
		\centering
		\includegraphics[width=\textwidth]{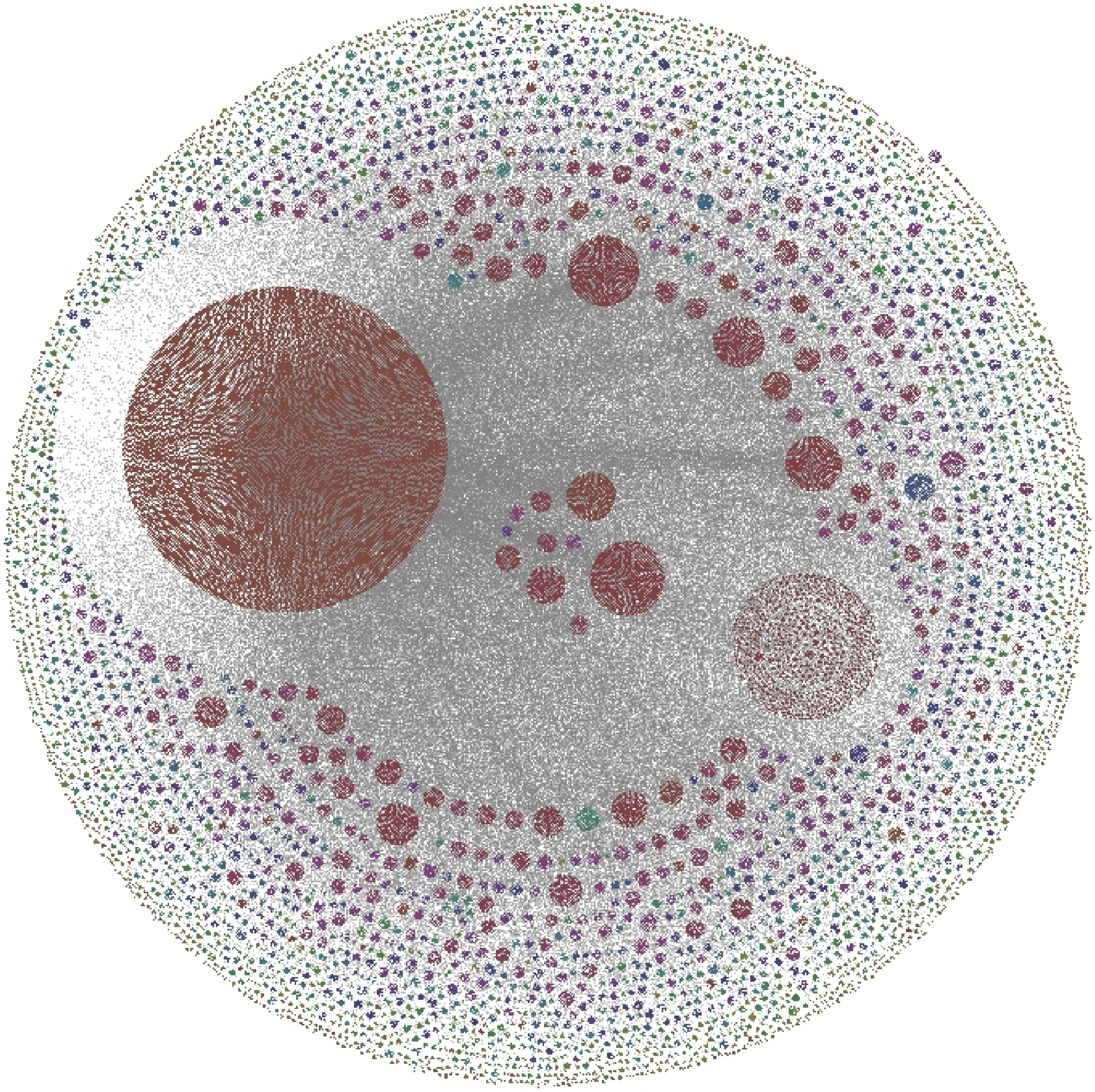}
		\caption{Best Result for Infomap approach}
		\label{fig:subfig20}
	}
	\end{subfigure}
	\begin{subfigure}{0.5\textwidth}
	{
		\centering
		\includegraphics[width=\textwidth]{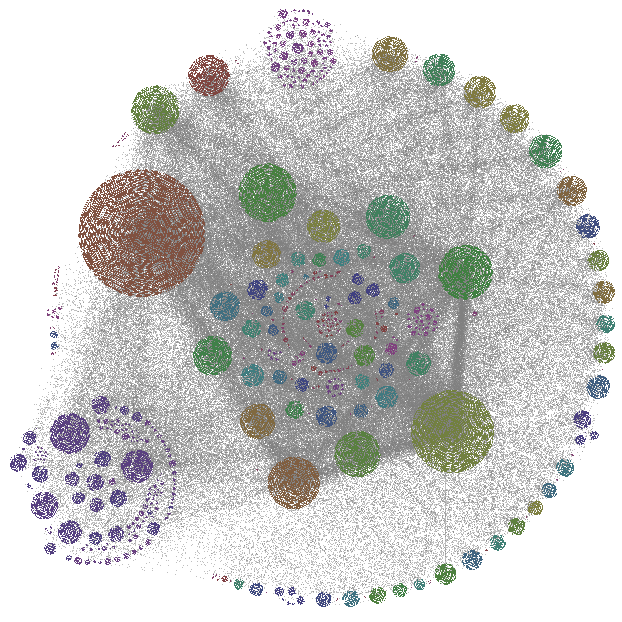}
		\caption{Best Result for OSLOM}
		\label{fig:subfig21}
	}
	\end{subfigure}
	\caption{Representing the $2$ best levels of hierarchy for Epn network w.r.t. the modularity ($Q$) criterion for various techniques.}
	\label{fig:fig8} 
\end{figure}

\begin{figure}[!ht]
	\begin{subfigure}{0.5\textwidth}
	{
		\centering
		\includegraphics[width=\textwidth]{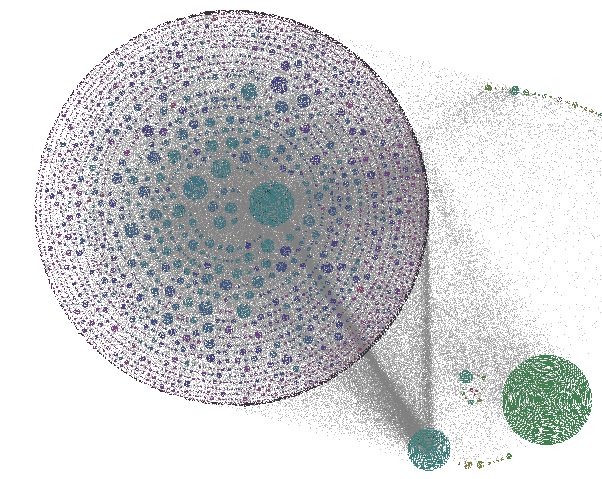}
		\caption{Best Result for MH-KSC algorithm}
		\label{fig:subfig18}
	}
	\end{subfigure}
		\begin{subfigure}{0.48\textwidth}
	{
		\centering
		\includegraphics[width=\textwidth]{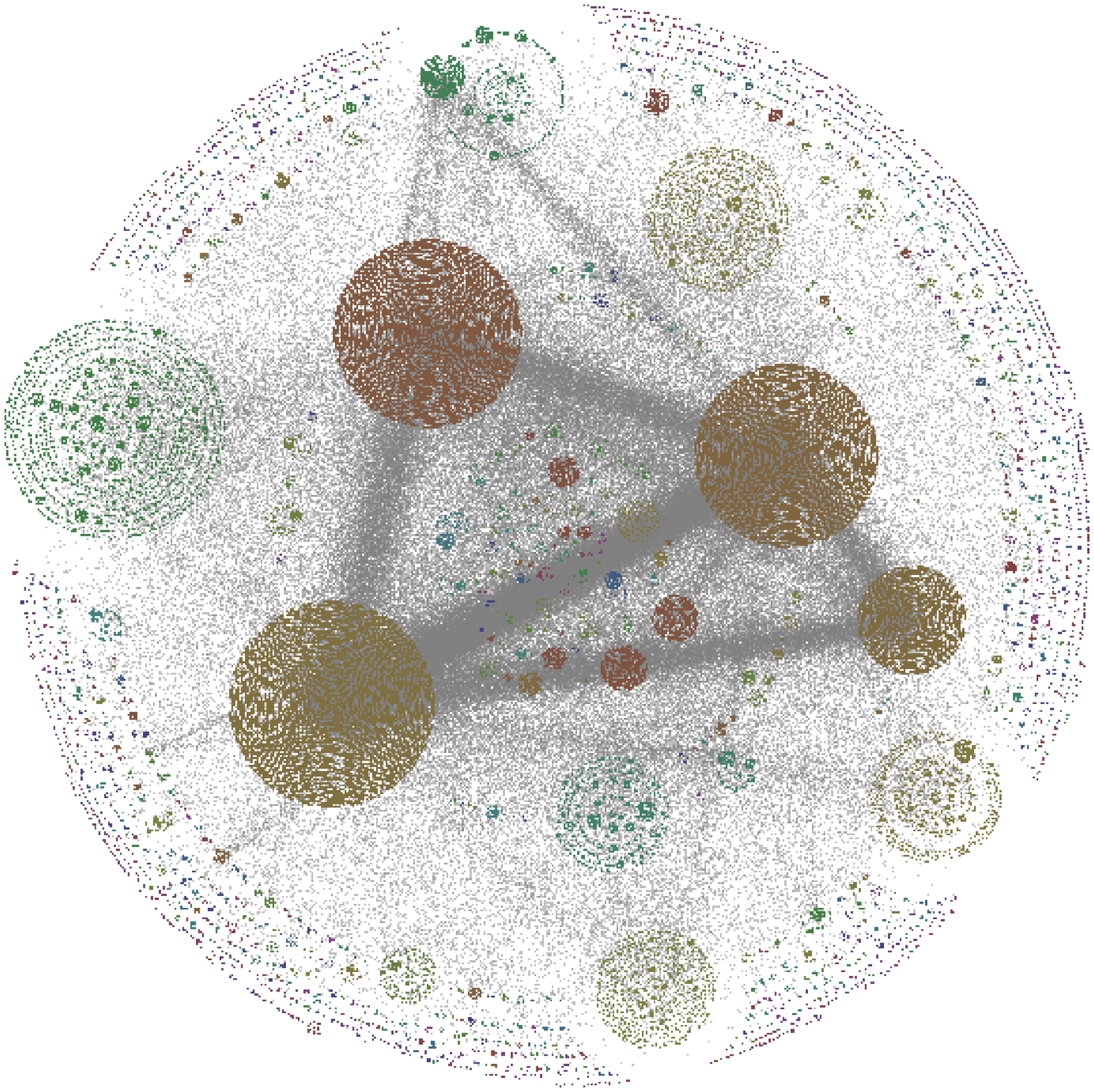}
		\caption{Best Result for Louvain method}
		\label{fig:subfig19}
	}
	\end{subfigure}
	\begin{subfigure}{0.5\textwidth}
	{
		\centering
		\includegraphics[width=\textwidth]{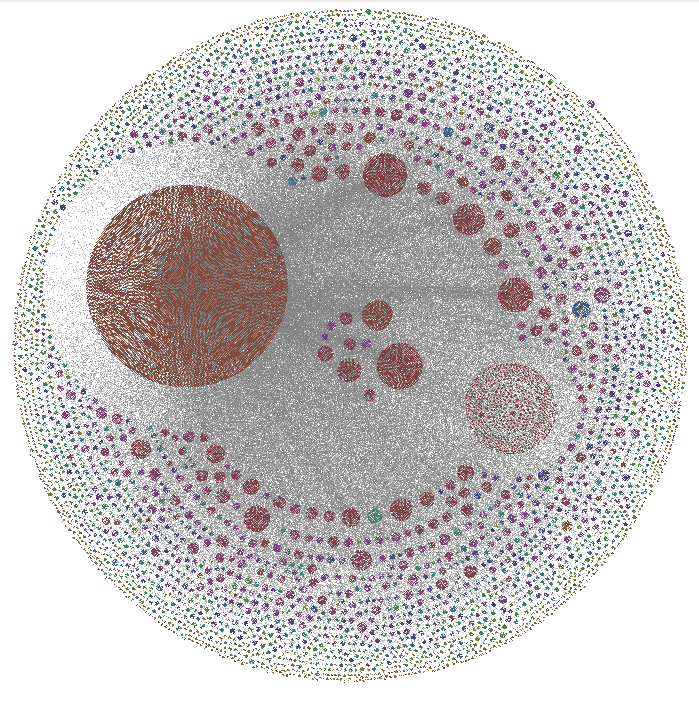}
		\caption{Best Result for Infomap approach}
		\label{fig:subfig20}
	}
	\end{subfigure}
	\begin{subfigure}{0.45\textwidth}
	{
		\centering
		\includegraphics[width=\textwidth]{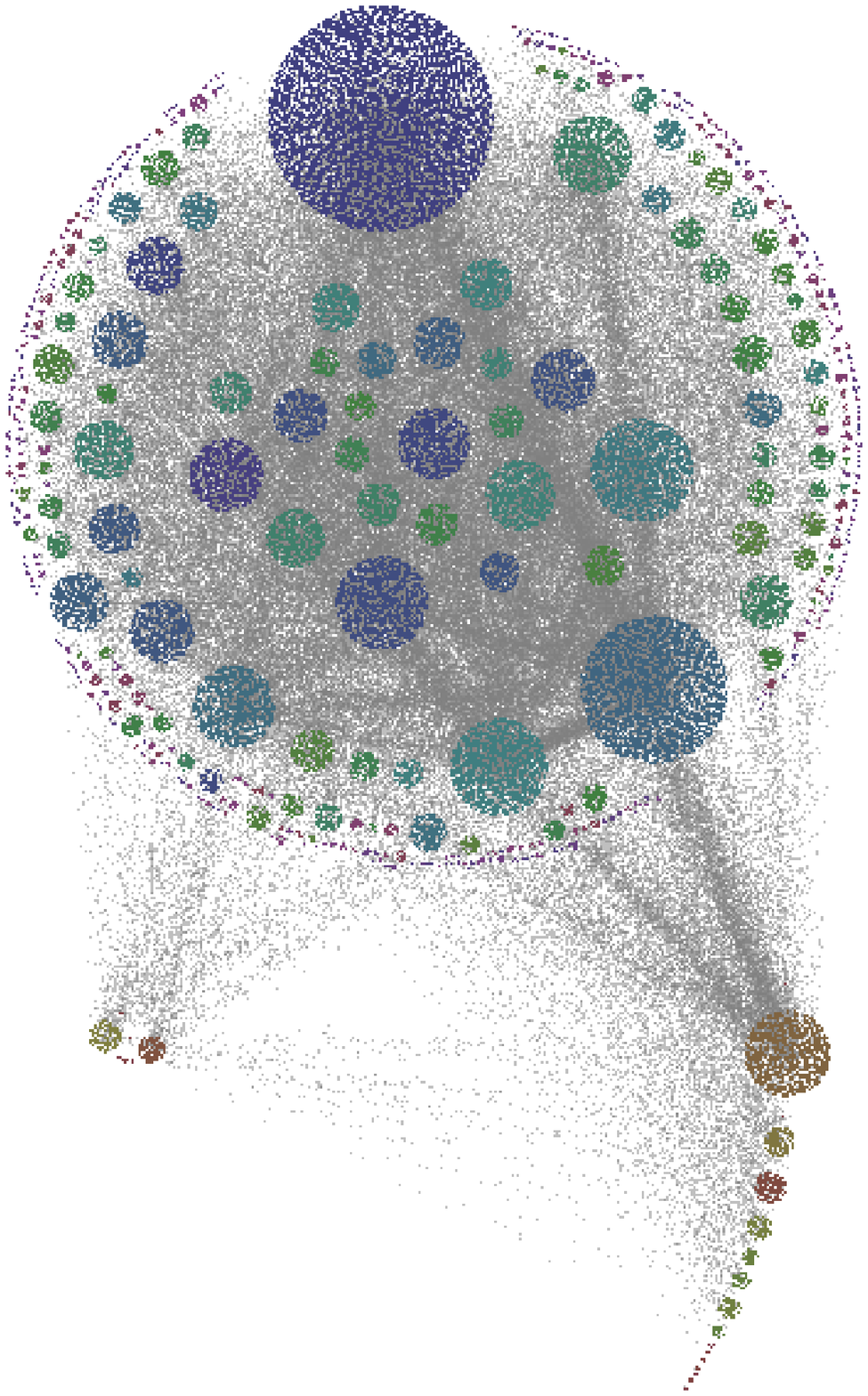}
		\caption{Best Result for OSLOM}
		\label{fig:subfig21}
	}
	\end{subfigure}
	\caption{Representing the $2$ best levels of hierarchy for Epn network w.r.t. the modularity ($CC$) criterion for various techniques.}
	\label{fig:fig9} 
\end{figure}

We provide a visualization of the $2$ best layers of hierarchy for Epn network based on the $Q$ and the $CC$ criterion for MH-KSC, Louvain, Infomap and OSLOM methods respectively in Figures \ref{fig:fig8} and \ref{fig:fig9}. The result for Infomap method in both the figures is the same as it only generates $2$ levels of hierarchy.
\section{Conclusion}\label{sec:sec5}
We proposed a new multilevel hierarchical kernel spectral clustering (MH-KSC) algorithm. The approach relies on the KSC primal-dual formulation and exploits the structure of the projections in the eigenspace. The projections of the validation set provided a set ( ${\cal T}$) of increasing distance thresholds. These distance thresholds were used along with affinity matrix obtained from the projections in an iterative procedure to obtain a multilevel hierarchical organization in a bottom-up fashion. We highlighted some of the necessary conditions for the feasibility of the approach to large scale networks. We showed that many real-life networks exhibit hierarchical structure. Our proposed approach was able to identify good quality clusters for both coarse as well as fine levels of granularity. We compared and evaluated our MH-KSC approach against several state-of-the-art large scale hierarchical community detection techniques.
\scriptsize{
\section*{Acknowledgements}
This work was supported by Research Council KUL: ERC AdG A-DATADRIVE-B, GOA/11/05 Ambiorics, GOA/10/09MaNet, CoE EF/05/006 Optimization in Engineering(OPTEC), IOF-SCORES4CHEM, several PhD/postdoc  and fellow grants; Flemish Government:FWO: PhD/postdoc grants, projects: G0226\-.06 (cooperative systems \& optimization), G0321.06 (Tensors), G.0302.07 (SVM/Kernel), G.0320.08 (convex MPC), G.0558.08 (Robust MHE), G.0557.08 (Glycemia2), G.0588.09 (Brain-machine) G.0377. 12 (structured models) research communities (WOG:ICCoS, ANMMM, MLDM); G.0377.09 (Mechatronics MPC) IWT: PhD Grants, Eureka-Flite$+$, SBO LeCoPro, SBO Climaqs, SBO POM, O\&O-Dsquare; Belgian Federal Science Policy Office: IUAP P6/04 (    DYSCO, Dynamical systems, control and optimization, 2007-2011); EU: ERNSI; FP7-HD-MPC (INFSO-ICT-223854), COST intelliCIS, FP7-EMBOCON (ICT-248940); Contract Research: AMINAL; Other:Helmholtz: viCERP, ACCM, Bauknecht, Hoerbiger. Johan Suykens is a professor at the KU Leuven, Belgium.
}   





\scriptsize{

}

\end{document}